\documentclass[aps,prd,showpacs,twocolumn,
superscriptaddress]{revtex4-1}

\usepackage{graphicx}
\usepackage{subfigure} 
\usepackage{dcolumn}

\graphicspath{{figures/}}
\usepackage{amsmath}
\usepackage{amssymb}
\usepackage{bm}
\usepackage{color}
\usepackage{comment}

\usepackage[normalem]{ulem}
\usepackage[dvipsnames]{xcolor}
\usepackage{hyperref}

\hypersetup{
colorlinks=true, % false: boxed links; true: colored links
linkcolor=blue,  % color of internal links
citecolor=cyan,  % color of links to bibliography
}

\begin{document}
\title{
Full waveform model for axisymmetric black hole mergers
}

\author{Song Li}
\email{leesong@shao.ac.cn}
\affiliation{Shanghai Astronomical Observatory, Shanghai, 200030, China }
\affiliation{School of Astronomy and Space Science, University of Chinese Academy of Sciences,
Beijing, 100049, China}

\author{Wen-Biao Han}
\email{wbhan@shao.ac.cn}
\affiliation{School of Fundamental Physics and Mathematical Sciences, Hangzhou Institute for Advanced Study, UCAS, Hangzhou 310024, China }
\affiliation{Shanghai Astronomical Observatory, Shanghai, 200030, China }
\affiliation{School of Astronomy and Space Science, University of Chinese Academy of Sciences,
Beijing, 100049, China}
\affiliation{International Centre for Theoretical Physics Asia-Pacific, Beijing/Hangzhou 100190, China}

\date{\today}
\begin{abstract}
In this work, we present a non-general relativity full waveform for general parametrization of axisymmetric black holes by extending our previous photon sphere+insipral model. Our model comprises two main components: an inspiral part obtained by using a phenomenological method in frequency-domain and a ringdown part derived from quasinormal modes associated with photon motion. For quantitatively revealing the influence of the deviation from Kerr black holes on the waveforms, we specify our model to the bumpy black holes, which are typical examples of non-general relativity black holes. The results show that the deviation from the Kerr quadrupole moment could be measured with a high accuracy. The new waveform model can be directly used to test black holes for the LIGO-Virgo-KAGRA observations, the third generation detectors and space-borne interferometers.
\end{abstract}

\maketitle

\section{Introduction}
In the past decade, there have been significant advancements in the study of gravitational waves (GWs). The detection of GW150914 by the LIGO-Virgo Collaboration in 2015 marked a major breakthrough as it was the first observation of gravitational wave events from a binary black hole system \cite{GW150914}. Since then, more compact binary systems have been discovered and reported by the LIGO-Virgo-KAGRA Collaboration. To date, over 90 compact binary systems have been identified \cite{GW_events_1, GW_events_2, GW_events_3, GW_events_4}, including two neutron star and black hole systems that are distinct from other binary black hole systems \cite{BN}. These unique compact binaries provide excellent scenarios for testing general relativity (GR) \cite{Test_GR_1, Test_GR_2, Test_GR_3, Test_GR_4, Test_GR_5, Test_GR_6}, gaining new insights into compact objects \cite{CO_1}, and potentially discovering new theories beyond GR. Ground-based detectors such as LIGO, Virgo, and KAGRA have opened up new avenues for understanding compact binary physics and astrophysics. Future space-based detectors such as Laser Interferometer Space Antenna (LISA), Taiji\cite{taiji}, and Tianqin\cite{tianqin} will offer different scenarios to improve our comprehension of the Universe further.

The no-hair theorem states that black holes in general relativity can be uniquely characterized by only a few fundamental properties, namely their mass, electric charge, and angular momentum. All other details, such as the matter that formed the black hole, are "hairless" and do not affect its external gravitational field. In 1915, a black hole model was introduced by Schwarzschild to characterize stationary and spherical spacetime. Subsequently, Kerr devised a broader black hole model, commonly referred to as the Kerr black hole, that possesses both stationary and axisymmetric properties. These two classical metrics, which are asymptotically flat, can be defined by two fundamental parameters: the mass denoted as $M$ and the spin represented by $a$. The no-hair theorem suggests that a black hole has information about its charge $q$, such as in the Reissner-Nordstrom black hole or the Kerr-Newman black hole. However, for real astrophysical black holes, their charge can usually be ignored. Gravity theories yield a wide range of metrics. Early metrics were Ricci-flat solutions of Einstein's field equations but had naked singularities or other pathologies \cite{Ricci_1, Ricci_2, Ricci_3}. Over recent years, researchers have studied Zipoy-Voorhees spacetime (also known as $\delta$-metric,$q$-metric, or $\gamma$-metric) \cite{ZV_1, ZV_2, ZV_3, ZV_4, ZV_5}, while for a more general metric solution, they have proposed and studied the $\delta$-Kerr metric \cite{deltaKerr_1,deltaKerr_2,deltaKerr_3}. This nonlinear superposition of the Zipoy-Voorhees metric with the Kerr metric represents a deformed Kerr solution. Based on the parametrized post-Newtonian(PPN) theory which describes strong gravity far away from sources, Konoplya, Rezzolla, and Zhidenko introduced the KRZ metric\cite{KRZ_16} to describe most general black hole spacetimes through finite adjustable quantities. The abundance of metrics derived from general relativity or alternative theories of gravity presents a fertile ground for investigating various astronomical phenomena in future research.

The Event Horizon Telescope (EHT) Collaboration recently captured shadow images of collapsed objects located at the center of two galaxies: M87* in an elliptical galaxy and Sagittarius-A* (SgrA*) in the Milky Way\cite{EHT_1, EHT_2, EHT_3, EHT_S}. These observations have sparked a new area of research focused on testing gravity theories and black hole solutions within a gravitational field regime that has never been tested before\cite{Test_GR_Sha_1, Test_GR_Sha_2, Test_GR_Sha_3}. This image is a result of light bending in the gravitational field of the source and has been extensively studied in GR\cite{Light_Bend_1, Light_Bend_2, Light_Bend_3, Light_Bend_4}. Theoretical explanations for these images can be found in references such as \cite{Image_1, Image_2, Image_3, Image_4, Image_5, Image_6, Image_7, Image_8, Image_9, Image_10}. The concept of the black hole shadow, which is associated with the photon sphere surrounding the central object, was explored in Ref. \cite{Claude_01}. Over several years, research on the photon sphere has included various approaches such as different images of black holes\cite{EHT_1, EHT_S, Lu_23}, ray-tracing codes\cite{Ray_1, Ray_2, Ray_3, Ray_4}, and accretion disks\cite{Acc_1, Acc_2}. These methods have provided effective ways to study black holes and general relativity. Several authors have used photon motion to investigate the black hole shadow using different metrics including the Kerr metric\cite{Li_14, Wei_19, Cunha_16}, the Kerr-Newman metric\cite {Vries_00}, and other rotating regular black hole metrics \cite {Ahmadjon_16, Li_21, Li_22}. Furthermore, the M87* shadow has been utilized by researchers to evaluate alternative theories of gravity such as superspinar \cite {Test_GR_Sha_3}and conformal massive gravity \cite {Jusufi_20}. McWilliams\cite{McWilliams_19} proposed a new waveform model for GR black holes called the backwards one-body (BOB) method that only considers photon motion without any phenomenological degrees of freedom.

The coalescence of two black holes involves three stages: inspiral, merger, and ringdown. During the inspiral phase, the black holes gradually approach each other. This phase refers to the few seconds before the merger. During the merger, they combine to form a single black hole. Finally, during the ringdown phase, the newly formed black hole undergoes oscillations, gradually releasing excess energy and angular momentum in the form of gravitational waves, until it reaches a stable state of equilibrium. Different phases have different methods to describe. In general, the inspiral phase is described by post-Newtonian theory while numerical relativity simulation provides a good description of the merger phase. Quasinormal modes are used to study the ringdown phase where there is still a slight perturbation in spacetime compared to a stable black hole. Black hole perturbation theory can be used to study the ringdown waveform. Well-known waveform models such as EOBNR and IMRPhenom rely on this idea \cite{Model_Nor_1, Model_Nor_2, Model_Nor_3, Model_Nor_4, Model_Nor_5, Model_Nor_6, Model_Nor_7, Model_Nor_8, Model_Nor_9, Model_Nor_10}. Currently, the Teukolsky equation is the most widely used method for calculating perturbations. This linear partial differential equation describes how scalar, vector, and gravitational perturbations of Kerr black holes evolve. By solving this equation with an outgoing boundary condition at infinity and an ingoing boundary condition at the horizon, complex frequencies known as quasinormal modes (QNMs) can be used to characterize gravitational waves. The QNMs are a spectrum of complex frequencies associated with gravitational waves. In addition to the fundamental QNMs, the overtone refers to higher-order vibrational modes that a black hole can exhibit during its ringdown phase. If we consider the ringdown phase early enough, before these overtones would have decayed it was shown that including them enhances waveform template accuracy, increases the signal-to-noise ratio (SNR), and allows more precise testing of the no-hair theorem\cite{Giesler_19, Iara_20}. Dhani \cite{Arnab_21} studies the insertion of negative-frequency modes (counterrotation), known as mirror modes, in addition to positive-frequency modes, known as regular modes, in the gravitational waveform. He shows that this will provide a more accurate description of the gravitational wave signal. Yang \textit{et al.}\cite{Yang_12} developed a method for calculating QNMs through the photon sphere. 

In this study, we proposed a full waveform model for the non-GR black hole by combining the inspiral and ringdown components. The inspiral component is constructed by the phenomenological method, while the ringdown component is obtained based on the photon sphere. We refer to this model as PSI-FD($\Psi_{\mathrm{FD}}$), which is an extension of our previous $\Psi$ (PSI: photon sphere + inspiral) model\cite{Psi_22}. The $\Psi$ model is a waveform model that consists of the inspiral part calibrated from post-Newtonian approximation and the ringdown part derived from the photon sphere. Reference~\cite{Psi_22} demonstrates the high accuracy of the $\Psi$ waveform model compared to numerical relativity (NR) waveforms.

This article is structured as follows: Sec.~\ref{KRZ} provides an introduction to the KRZ metric and highlights key considerations when using this metric. We then combine the KRZ metric with the bumpy black hole metric to establish the relationship between the parameters of the KRZ metric and the quadrupole moment. In Sec.~\ref{Inspiral}, we present a methodology for extracting the inspiral waveform from photon motion. Subsequently, Sec.~\ref{Ringdown} outlines the derivation of QNMs through the photon sphere and describes how to obtain the ringdown waveform using these QNMs. Finally, we combine the inspiral and ringdown waveform to obtain the complete waveform. Our results and findings are presented in Sec.~\ref{Conclusion}. We have fixed units such that $G=c=1$.

\section{Parametrized Black Hole metric and the application}\label{KRZ}
The KRZ metric, constructed by Konoplya, Rezzolla, and Zhidenko~\cite{KRZ_16} proposes a parametrization for general stationary and axisymmetric black holes. They develop a model-independent framework that parametrizes the most generic black hole geometry by a finite number of tunable quantities. By adjusting these quantities, a number of famous black hole metrics, such as the Kerr metric, are found exactly and in the whole space. The metric with this form:
\begin{eqnarray}\label{metric}
d s^{2}&=&-\frac{N^{2}(\tilde{r}, \theta)-W^{2}(\tilde{r}, \theta) \sin ^{2} \theta}{K^{2}(\tilde{r}, \theta)} d t^{2}\nonumber\\&&-2 W(\tilde{r}, \theta) \tilde{r} \sin ^{2} \theta d t d \phi  \nonumber\\&&+K^{2}(\tilde{r}, \theta) \tilde{r}^{2} \sin ^{2} \theta d \phi^{2}\nonumber\\&&+\Sigma(\tilde{r}, \theta)\left(\frac{B^{2}(\tilde{r}, \theta)}{N^{2}(\tilde{r}, \theta)} d \tilde{r}^{2}+\tilde{r}^{2}d \theta^{2}\right), 
\end{eqnarray}

where $\tilde{r}=r/M, \ \tilde{a}=a / M$ and the other metric functions are defined as

\begin{eqnarray}
\Sigma&=&1+a^{2} \cos ^{2} \theta/\tilde{r}^{2}\ ,\\
{N}^{2}&=&\left( 1-{{r}_{0}}/\tilde{r} \right)\nonumber\\&&\left[ 1-{{\epsilon }_{0}}{{r}_{0}}/\tilde{r}+\left( {{k}_{00}}-{{\epsilon }_{0}} \right)r_{0}^{2}/{{{\tilde{r}}}^{2}}+{{\delta }_{1}}r_{0}^{3}/{{{\tilde{r}}}^{3}} \right]\nonumber\\&&+[ {{a}_{20}}r_{0}^{3}/{{{\tilde{r}}}^{3}}+{{a}_{21}}r_{0}^{4}/{{{\tilde{r}}}^{4}}+{{k}_{21}}r_{0}^{3}/{{{\tilde{r}}}^{3}}L]{{\cos }^{2}}\theta\ ,\\
B&=&1+\delta_{4} r_{0}^{2} / \tilde{r}^{2}+\delta_{5} r_{0}^{2} \cos ^{2} \theta / \tilde{r}^{2}\ ,\\
W&=&\left[w_{00} r_{0}^{2} / \tilde{r}^{2}+\delta_{2} r_{0}^{3} / \tilde{r}^{3}+\delta_{3} r_{0}^{3} / \tilde{r}^{3} \cos ^{2} \theta\right] / \Sigma\ ,\\
K^{2}&=&1+a W / r\nonumber\\&&+\left\{k_{00} r_{0}^{2} / \tilde{r}^{2}+k_{21} r_{0}^{3} / \tilde{r}^{3}L \cos ^{2} \theta\right\} / \Sigma\ ,
\\ \label{L}
L&=&\left[1+\frac{k_{22}\left(1-r_{0} / \tilde{r}\right)}{1+k_{23}\left(1-r_{0} / \tilde{r}\right)}\right]^{-1}\ .
\end{eqnarray}

This paper adopts the parameters defined as follows:

\begin{eqnarray}
a_{20}&=&2 \tilde{a}^{2} / r_{0}^{3}, \\ 
a_{21} &=&-\tilde{a}^{4} / r_{0}^{4}+\delta_{6}, \\ \epsilon_{0}&=&\left(2-r_{0}\right) / r_{0}, \\ k_{00}&=&\tilde{a}^{2} / r_{0}^{2} ,\\ 
k_{21}&=&\tilde{a}^{4} / r_{0}^{4}-2 \tilde{a}^{2} / r_{0}^{3}-\delta_{6}, \\ 
w_{00}&=& 2 \tilde{a} / r_{0}^{2}, \\ 
k_{22}&=&-\tilde{a}^{2} / r_{0}^{2}+\delta_{7}, \\ k_{23}&=&\tilde{a}^{2} / r_{0}^{2}+\delta_{8},
\end{eqnarray}

The dimensionless parameter $\delta_{i}$, where $i=1,2,3,4,5,6,7,8$, describes the deformation of various parameters in metric~(\ref{metric}). Specifically, $g_{tt}$ is deformed by $\delta_{1}$, while $\delta_{2}$ and $\delta_{3}$ correspond to spin deformations. Additionally, $\delta_{4}$ and $\delta_{5}$ relate to deformations of $g_{rr}$, and $\delta_6$ is for event horizon deformation. When all values of $\delta_i$ are zero ($\delta_i = 0$), the KRZ metric reduces to the Kerr metric specified in~(\ref{metric}). Furthermore, if $\tilde{a}=0$, it reduces to the Schwarzschild metric.

The parameter $r_0$ represents the equatorial radius of the event horizon. Some papers have utilized the KRZ metric to explore the general parametrization of axisymmetric black holes. However, these papers erroneously employed the definition ${{r}_{0}}=M+\sqrt{{{M}^{2}}-{{a}^{2}}}$, which is only valid in the context of the Kerr metric. We provide two examples to illustrate this point.

\subsubsection{EDGB black hole metric}
The Einstein-dilaton-Gauss-Bonnet (EDGB) gravity is a modified theory of gravity that extends general relativity (GR) by including an additional scalar field (dilaton) and the Gauss-Bonnet term, a curvature term that arises from higher-dimensional theories of gravity. The authors of this paper\cite{KRZ_16} provided the parametrization for an EDGB black hole:

\begin{equation}
r_{0}  =  2 M\left(1-\frac{\chi^{2}}{4}-\frac{49 \zeta}{80}+\frac{128171 \chi^{2} \zeta}{588000}\right)+\mathcal{O}\left(\chi^{4}, \zeta^{2}\right)\label{r0_EDGB}
\end{equation}

\begin{eqnarray}
\delta_{1}&=&-\frac{17 \zeta}{60}\left(1-\frac{324899 \chi^{2}}{166600}\right)+\mathcal{O}\left(\chi^{4}, \zeta^{2}\right),\\
\delta_{2}&=&-\frac{63 \chi \zeta}{160}+\mathcal{O}\left(\chi^{3}, \zeta^{2}\right),\\
\delta_{3}&=&\mathcal{O}\left(\chi^{3}, \zeta^{2}\right),\\
\delta_{4}&=&-\frac{361 \zeta}{240}\left(1-\frac{51659 \chi^{2}}{176890}\right)+\mathcal{O}\left(\chi^{4}, \zeta^{2}\right),\\
\delta_{5}&=&\frac{175629}{196000} \chi^{2} \zeta+\mathcal{O}\left(\chi^{4}, \zeta^{2}\right).
\end{eqnarray}

\begin{equation}
\chi \equiv \frac{a}{M}=\frac{J}{M^2} ,\zeta \equiv \frac{16\pi \alpha^2}{\beta M^4} 
\end{equation}
where $\alpha$ and $\beta$ are two coupling constants in Einstein-dilaton-Gauss-bonnet theory,
$\alpha$ represents the coupling of higher curvature, while $\beta$ accounts for the coupling with the scalar field.

\subsubsection{Dilaton black hole metric}
Additionally, the authors provided the parametrization for a dilaton black hole:

\begin{equation}
r_{0} =\sqrt{(\mu+b+\sqrt{\mu^2-a^2} )^2-b^2} \label{r0_Dilaton}
\end{equation}

\begin{eqnarray}
\delta_{1}&=& \frac{2(\mu+b)\left[2 b^{2}+r_{0}^{2}+\left(2 r_{0}-3 b\right) \sqrt{r_{0}^{2}+b^{2}}\right]}{r_{0}^{2} \sqrt{r_{0}^{2}+b^{2}}} \nonumber\\&&-3 \frac{r_{0}^{2}+a^{2}}{r_{0}^{2}},\\
\delta_{2}&=&\frac{2 a(\mu+b)\left(b+r_{0}-\sqrt{r_{0}^{2}+b^{2}}\right)}{r_{0}^{3}},\\
\delta_{3}&=&0,\\
\delta_{4}&=&\frac{r_{0}}{\sqrt{r_{0}^{2}+b^{2}}}-1,\\
\delta_{5}&=&0,
\end{eqnarray}
where $\mu$ and $b$ are the dilaton parameters.

Equations (\ref{r0_EDGB}) and (\ref{r0_Dilaton}) demonstrate that when considering specific metrics, such as the EDGB or Dilaton metrics, the equatorial radius of the event horizon ($r_0$) is not equal to ${{r}_{0}}=M+\sqrt{{{M}^{2}}-{{a}^{2}}}$. Therefore, when utilizing the KRZ metric to obtain the value of $r_0$, it is crucial to focus on the appropriate metric. This concept is easily comprehensible because the KRZ metric is a general metric. However, if the value of $r_0$ is fixed using the definition ${{r}_{0}}=M+\sqrt{{{M}^{2}}-{{a}^{2}}}$, then the universality of the metric is lost.

\subsection{Bumpy black hole}
General relativity predicts the existence of compact objects known as black holes, whose spacetimes are solely determined by their mass, spin, and charge in vacuum, in accordance with the ``no-hair" theorem. Collins and Hughes\cite{Bumpy_BH_1} proposed the existence of an exception, called bumpy black holes. These objects possess a multipolar structure closely resembling that of black holes but with some deviation. When the deviation is set to zero, bumpy black holes reduce to standard black holes such as the Schwarzschild black hole or the Kerr black hole. The bumpy Kerr black hole metric can be expressed in the Boyer-Lindquist coordinates, as shown below:

\begin{equation}
\begin{aligned}
d s^2= & -e^{2 \psi_1}\left(1-\frac{2 M r}{\Sigma}\right) d t^2+\\
&e^{2 \psi_1-\gamma_1}\left(1-e^{\gamma_1}\right) \frac{4 a^2 M r \sin ^2 \theta}{\Delta \Sigma} d t d r \\
&-e^{2 \psi_1-\gamma_1} \frac{4 a M r \sin ^2 \theta}{\Sigma} d t d \phi +e^{2 \gamma_1-2 \psi_1}\left(1-\frac{2 M r}{\Sigma}\right)^{-1}\\
&[1+e^{-2 \gamma_1}\left(1-2 e^{\gamma_1}\right) \frac{a^2 \sin ^2 \theta}{\Delta}-\\
&e^{4 \psi_1-4 \gamma_1}\left(1-e^{\gamma_1}\right)^2 \frac{4 a^4 M^2 r^2 \sin ^4 \theta}{\Delta^2 \Sigma^2}] d r^2 \\
& -2\left(1-e^{\gamma_1}\right) a \sin ^2 \theta[e^{-2 \psi_1}\left(1-\frac{2 M r}{\Sigma}\right)^{-1}-\\
&e^{2 \psi_1-2 \gamma_1} \frac{4 a^2 M^2 r^2 \sin ^2 \theta}{\Delta \Sigma(\Sigma-2 M r)}] d r d \phi \\
& +e^{2 \gamma_1-2 \psi_1} \Sigma d \theta^2+\Delta [e^{-2 \psi_1}\left(1-\frac{2 M r}{\Sigma}\right)^{-1}-\\
&e^{2 \psi_1-2 \gamma_1} \frac{4 a^2 M^2 r^2 \sin ^2 \theta}{\Delta \Sigma(\Sigma-2 M r)}] \sin ^2 \theta d \phi^2
\end{aligned}
\end{equation}

The bumpy Kerr black hole metric can be expressed in the form $g_{\alpha\beta}=g^{\mathrm{Kerr}}_{\alpha\beta}+b_{\alpha\beta}$, where $g^{\mathrm{Kerr}}_{\alpha\beta}$ denotes the Kerr metric. In the above equation, $\Delta\equiv r^2-2Mr+a^2$, and $\gamma_1$ and $\psi_1$ denote the perturbation potentials arising from the mass moment and spin moment perturbations, respectively. The definitions of $\gamma_1$ and $\phi_1$ are detailed in \cite{Bumpy_BH_1, Bumpy_BH_2}. The bumpy Kerr black hole metric reduces to the Kerr black hole metric, i.e., $\gamma_1=\phi_1=0$, in the absence of perturbations.

\begin{eqnarray}
b_{t t}&=&-2\left(1-\frac{2 M r}{\Sigma}\right) \psi_1, \\
b_{t r}&=&-\gamma_1 \frac{2 a^2 M r \sin ^2 \theta}{\Delta \Sigma}, \\
b_{t \phi}&=&\left(\gamma_1-2 \psi_1\right) \frac{2 a M r \sin ^2 \theta}{\Sigma}, \\
b_{r r}&=&2\left(\gamma_1-\psi_1\right) \frac{\Sigma}{\Delta} \\
b_{r \phi}&=&\gamma_1\left[\left(1-\frac{2 M r}{\Sigma}\right)^{-1}-\frac{4 a^2 M^2 r^2 \sin ^2 \theta}{\Delta \Sigma(\Sigma-2 M r)}\right] a \sin ^2 \theta, \\
b_{\theta \theta}&=&2\left(\gamma_1-\psi_1\right) \Sigma \\
b_{\phi \phi}&=&\left[\left(\gamma_1-\psi_1\right) \frac{8 a^2 M^2 r^2 \sin ^2 \theta}{\Delta \Sigma(\Sigma-2 M r)}-2 \psi_1\left(1-\frac{2 M r}{\Sigma}\right)^{-1}\right]\nonumber\\&&
\Delta \sin ^2 \theta .
\end{eqnarray}

Vigeland and Hughes\cite{Bumpy_BH_2} proposed the quadrupole bumps(i.e., $l=2$) in the Boyer-Lindquist coordinates:

\begin{equation}
\begin{aligned}
\psi_1^{l=2}(r, \theta) & =\frac{B_2 M^3}{4} \sqrt{\frac{5}{\pi}} \frac{1}{d(r, \theta, a)^3}\left[\frac{3 L(r, \theta, a)^2 \cos ^2 \theta}{d(r, \theta, a)^2}-1\right], \\
\gamma_1^{l=2}(r, \theta) & =B_2 \sqrt{\frac{5}{\pi}}\\
[\frac{L(r, \theta, a)}{2} &\frac{\left[c_{20}(r, a)+c_{22}(r, a) \cos ^2 \theta+c_{24}(r, a) \cos ^4 \theta\right]}{d(r, \theta, a)^5}-1],\label{B2_first}
\end{aligned}
\end{equation}
where

\begin{equation}
\begin{aligned}
d(r, \theta, a)&=\sqrt{r^2-2 M r+\left(M^2+a^2\right) \cos ^2 \theta} \\
L(r, \theta, a)&=\sqrt{(r-M)^2+a^2 \cos ^2 \theta} \\
\end{aligned}
\end{equation}
and
\begin{equation}
\begin{aligned}
c_{20}(r, a)&=2(r-M)^4-5 M^2(r-M)^2+3 M^4, \\
c_{22}(r, a)&=5 M^2(r-M)^2-3 M^4+a^2\left[4(r-M)^2-5 M^2\right], \\
c_{24}(r, a)&=a^2\left(2 a^2+5 M^2\right) .
\end{aligned}
\end{equation}

By selecting the appropriate parameters, the KRZ metric can be reduced to the bumpy black hole metric. Since the KRZ metric does not provide an exact value for the black hole's quadrupole moment, we aim to utilize the quadrupole moment in the bumpy black hole metric to correspond to $\delta_i$. Upon performing these calculations, we determined that selecting specific values for $\delta_i$ results in the reduction of the KRZ metric to the bumpy black hole metric:

\begin{equation}
\begin{aligned}
\delta_1 =& \left \{[(2\psi_1+1)(1-\frac{2M}{r})](1-\delta_6\frac{r_0^3}{r^3}\mathrm{cos^2\theta})-(1-\frac{r_0}{r} )  \right \}\\
&/[\frac{r_0^3}{r^3}(1-\frac{r_0}{r}  ) ],\\
\delta_6 = &\frac{2\psi_1r^5}{r_0^3}\mathrm{tan^2\theta},\\
\delta_2 =&\delta_3=\delta_4=\delta_5=0.
\end{aligned}
\end{equation}

\begin{figure*}
\centering
\includegraphics[width=0.45 \textwidth]{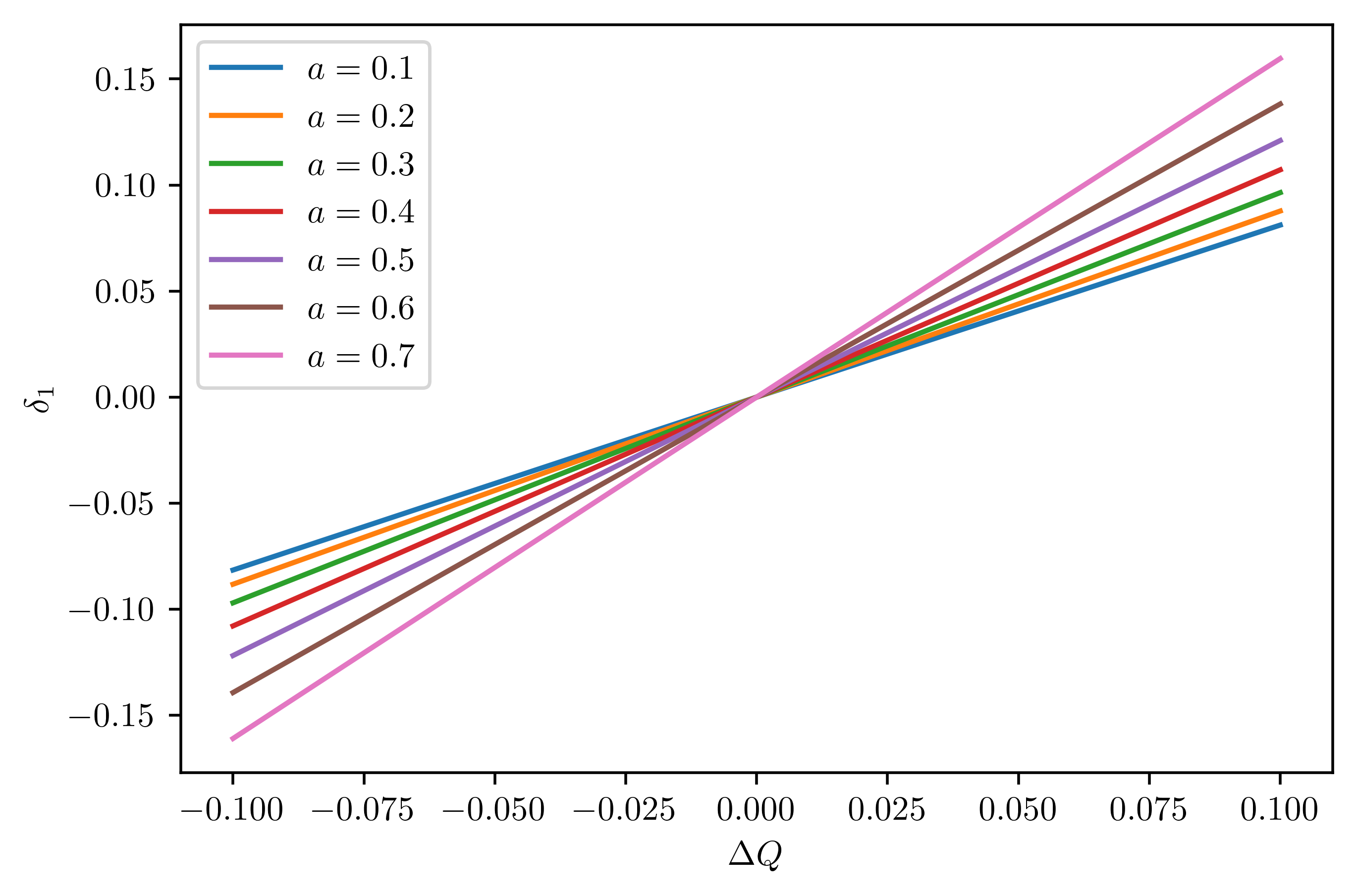}
\includegraphics[width=0.45 \textwidth]{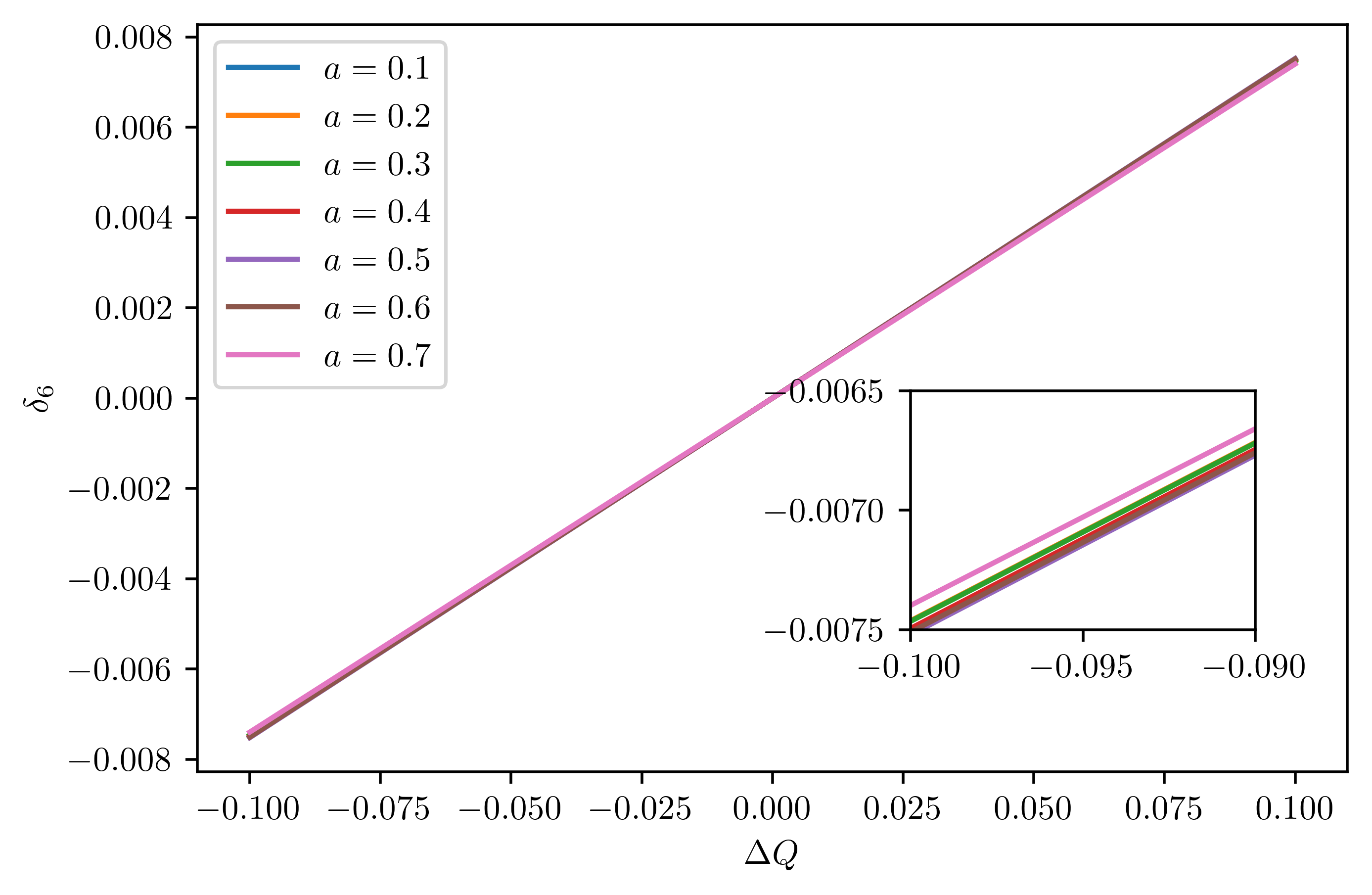}
\caption{The relationship between the $\Delta Q$ and $\delta_1(\delta_6)$ in the bumpy black hole metric with different spin $a$ from 0.1 to 0.7. The left panel shows the relationship between the $\Delta Q$ and $\delta_1$. The right panel shows the relationship between the $\Delta Q$ and $\delta_6$.}\label{DeltaQ_delta_Bumpy}
\end{figure*}

The quadrupole moment is given by the following equation:
\begin{equation}
{Q}=-Ma^2-B_2M^3\sqrt{5/4\pi}={Q}_K+\Delta{Q} 
\end{equation}
where $B_2$ is the parameter that appears in Eq.~(\ref{B2_first}). Figur/e ~\ref{DeltaQ_delta_Bumpy} displays the relationship between $\delta_1$, $\delta_6$, and $\Delta{Q}$ with different spins $a$ ranging from 0.1 to 0.7. The left panel displays the values of $\delta_1$ for different spins and $\Delta Q$ values. A larger $\Delta Q$ corresponds to a lower spin for the same value of $\delta_1$, suggesting that $\delta_1$ has a greater impact on the quadrupole moment at lower values of spin. The right panel illustrates that the influence of spin on $\delta_6$ is negligible. Additionally, the figure reveals that $\Delta Q$ increases with increasing values of $\delta_1$ and $\delta_6$, which represent the deviation of the Kerr metric. Furthermore, the increase in $\Delta Q$ is more pronounced with increasing $\delta_6$.

\section{The inspiral waveform in KRZ Black Holes}\label{Inspiral}

The inspiral phase waveform can be calculated through the geodesic motion, and we will focus on the deformation parameter $\delta_1$ in this section. For simplicity, we assume that all other deformation parameters are zero when considering one single parameter. In the following section, we will provide a brief overview of the derivation for the deformation phase, with details available in the paper\cite{KRZ_ins}. In Sec.~\ref{KRZ}, we have demonstrated that this value of ${{r}_{0}}=M+\sqrt{{{M}^{2}}-{{a}^{2}}}$ is only applicable in the Kerr metric. Therefore, $r_0=$ needs to be determined for varied non-GR black holes. As an example, we concentrate on a specific metric, namely the bumpy black hole metric, to determine the corresponding horizon radius $r_0$ and the deformation phase. To gain a deeper understanding, we investigate the quadrupole moment $Q$ rather than the deformation parameters $\delta_i$. The normalization of the four-velocity requires that $u^{\mu}u_{\mu}=-1$. To simplify the equation, we choose $\dot{\theta}$ equal to zero. Solving the equation(i.e, $u^{\mu}u_{\mu}=-1$) yields the following result:

\begin{equation}
V_{\mathrm{eff}}=g_{rr}\dot{r}^2=-1-g_{tt}\dot{t}^2-g_{\phi\phi}\dot{\phi}^2,
\end{equation}

To simplify the above equation, we can use the specific energy(the energy per unit mass) and specific angular momentum(the angular momentum per unit mass) of a particle:
\begin{eqnarray}
E=-(g_{tt}u^{t} +g_{t\phi }u^{\phi}),\label{energy}\\ 
L=g_{\phi t}u^{t} +g_{\phi\phi }u^{\phi}, \label{momentum}
\end{eqnarray}
and they are constants because the KRZ spacetime is stationary and axisymmetric. Through Eq.~(\ref{energy}) and (\ref{momentum}), we can get the expressions of $\dot{t}$ and $\dot{\phi}$:
\begin{eqnarray}
\dot{t}=-\frac{Lg_{t\phi}+Eg_{\phi\phi}}{g_{tt}g_{\phi\phi}-g_{t\phi}^2}\label{ut} ,\\ 
\dot{\phi}=\frac{Lg_{tt}+Eg_{t\phi}}{g_{tt}g_{\phi\phi}-g_{t\phi}^2}  . \label{uphi}
\end{eqnarray}

and then with Eq.~(\ref{ut}) and (\ref{uphi}) we can rewrite $V_{\mathrm{eff}}$ in terms of $E$, $L$ as

\begin{equation}
\begin{aligned}
V_{\mathrm{eff}}= & -1+E^2+\frac{2 M}{r}+\frac{L^2(2 M-r)}{r^3}+\frac{8 \delta_1 M^3(2 M-r)}{r^4} \\
& +\frac{8 \delta_1 L^2 M^3(2 M-r)}{r^6}+\mathcal{O}\left[\delta_1^2\right] .
\end{aligned}
\end{equation}

The determination of the energy and angular momentum of circular orbits relies on the condition that $V_{\mathrm{eff}}=dV_{\mathrm{eff}}/dr=0$. By satisfying this condition, it is possible to express the energy and angular momentum as the sum of the GR term and a small perturbation that relates to the deformation parameter $\delta_1$:
\begin{equation}
\begin{aligned}
E=E^{GR}+\delta E,\\
L=L^{GR}+\delta L.
\end{aligned}
\end{equation}
where
\begin{eqnarray}
\begin{aligned}
E^{\mathrm{GR}} & & = & \sqrt{\frac{4 M^{2}-4 M r+r^{2}}{(r-3 M) r}}, \\
L^{\mathrm{GR}} & & = & \sqrt{\frac{M r^{2}}{r-3 M}},
\end{aligned}
\end{eqnarray}

\begin{eqnarray}
\begin{aligned}
\delta E & =-\frac{2 M^{3}(r-2 M)}{r^{5 / 2}(r-3 M)^{3 / 2}} \delta_{1}+\mathcal{O}\left[\delta_{1}^{2}\right], \\
\delta L & =-\frac{6 M^{5 / 2}(r-2 M)^{2}}{r^{2}(r-3 M)^{3 / 2}} \delta_{1}+\mathcal{O}\left[\delta_{1}^{2}\right].
\end{aligned}
\end{eqnarray}

By considering the far-field limit in which $L=r^2\dot{\phi}\to r^2\Omega$ (where $\Omega=d\phi/dt$ refers to the angular velocity of the body as observed by a distant observer), one can obtain the following result:

\begin{equation}
\Omega^2=\frac{M}{r^3}\left[1+\frac{3 M}{r}+\frac{9 M^2}{r^2}-\frac{12 M^2}{r^2} \delta_1+\mathcal{O}\left(\delta_1^2, \frac{M^3}{r^3}\right)\right] .
\end{equation}

For circular orbits, it is possible to express the system's total energy ($E_T$) as the effective energy of a single body in the rest frame of the other.
\begin{equation}
E_T = m+E_b = m[1+2\eta(E_{\mathrm{eff}}-1)],
\end{equation}

The parameters in the equation include the binding energy, $E_b$, the symmetric mass ratio, $\eta=\mu/m$, and the reduced mass, $\mu$, which is defined as $\mu=m_1m_2/m$ ($m=m_1+m_2$, where $m_1$ and $m_2$ are the masses of the two bodies),

\begin{equation}
E_{\mathrm{eff}}=g_{tt}(1+\frac{L^2}{r^2})^{1/2},
\end{equation}

We separate the rest-mass energy $m$ from the binding energy $E_b$ to express the latter as a sum of its general relativity term and a correction:
\begin{equation}
E_{\mathrm{b}}=E_{\mathrm{b}}^{\mathrm{GR}}-\frac{\eta m^{2}}{2 r}\left[4 \delta_{1}\left(\frac{m}{r}\right)^{2}+\mathcal{O}\left(\delta_{1}{ }^{2}, \frac{m^{3}}{r^{3}}\right)\right]
\end{equation}

To simplify calculations, it is possible to express the binding energy, $E_b$, as a function of the orbital frequency, $\nu=\Omega/2\pi$:
\begin{equation}
\frac{E_{\mathrm{b}}(\nu)}{\mu}=\frac{E_{\mathrm{b}}^{\mathrm{GR}}(\nu)}{\mu}-4 \delta_{1}(2 \pi m \nu)^{2}+\mathcal{O}\left[\delta_{1}^{2},(2 \pi m \nu)^{8 / 3}\right] .
\end{equation}

The orbital phase is given by
\begin{equation}
\phi(\nu)=\int^{\nu} \Omega d t=\int^{\nu} \frac{1}{\dot{E}}\left(\frac{d E}{d \Omega}\right) \Omega d \Omega,
\end{equation}
the variable $\dot E$ describes the speed at which the binding energy is altered due to the emission of gravitational waves, which comprises two parts: the conservative sector and the dissipative sector. Our focus is primarily on the conservative sector of the gravitational waves emission, and the dissipative sector is assumed to be unaffected. Based on this study\cite{KRZ_ins}, we only need to employ the quadrupole formulation up to the post-Newtonian order of 0PN for determining the modification in binding energy,

\begin{equation}
\dot E_\mathrm{GR}^{0\mathrm{PN}}=-\frac{32}{5}\eta^2m^2r^4\Omega^6  
\end{equation}

Then we can get the expression for the orbital phase evolution:
\begin{equation}
\phi(\nu )=\phi^{\mathrm{0PN}}_{\mathrm{GR}}(\nu )-\frac{25}{e\eta}(2\pi m\nu)^{-\frac{1}{3} }\delta_1+\mathcal{O}[\delta_1^2]
\end{equation}
where
\begin{equation}
\phi^{\mathrm{0PN}}_{\mathrm{GR}}(\nu )=-\frac{1}{32\eta}(2\pi m\nu)^{-5/3}.
\end{equation}

In the stationary phase approximation, the Fourier transform of $\phi$ is given by $\Phi^{\mathrm{GR}}_{\mathrm{GW}}(f)=2\phi(t_0)-2\pi ft_0$, where $t_0$ is the stationary time,  $\nu(t_0)=f/2$ and $f$ is the Fourier frequency. Then we can get

\begin{equation}
\Psi_{\mathrm{GW} }(f)=\Psi^{\mathrm{GR,0PN} }_{\mathrm{GW} }(f)-\frac{75}{8}u^{-1/3}\eta^{-4/5}\delta_1+\mathcal{O}[\delta_1^2],
\end{equation}
where
\begin{equation}
\Psi^{\mathrm{GR,0PN} }_{\mathrm{GW} }(f)=-\frac{3u^{-5/3}}{128},
\end{equation}
and $u=\eta \pi mf$.

So the deformation of $\delta_1$ on the phase can be expressed as
\begin{equation}
\phi^{\delta_1}_{\mathrm{KRZ}} = -\frac{75}{8}u^{-1/3}\eta^{-4/5}\delta_1
\end{equation}

We can apply the same method for determining the deformation parameters besides the one already discussed.

Deformation parameter $\delta_2$:
\begin{equation}
\phi^{\delta_2}_{\mathrm{KRZ}} = -\frac{85}{3\eta}[1+\mathrm{log}(u)]\delta_2.
\end{equation}

Deformation parameter $\delta_3$:
\begin{equation}
\phi^{\delta_3}_{\mathrm{KRZ}} = -C_1\delta_3\mathrm{cos^2}(\theta)f^{5/3},
\end{equation}
$C_1 = 48\cdot2^{-1/3}m^{5/3}/(\eta \pi)$.

Deformation parameter $\delta_6$:
\begin{equation}
\phi^{\delta_6}_{\mathrm{KRZ}} = -C_2\delta_6\mathrm{cos^2}(\theta)f^{-1},
\end{equation}
$C_2 = 40m/\eta$.

The deformation parameters $\delta_4$ and $\delta_5$ have no effect on the motion of geodesics and therefore they cannot be constrained within the framework. 

Having obtained the phase deformation using the deformation parameters $\delta_i$, we can apply it to a more accurate model such as the PhenomD model\cite{phenomD}. The PhenomD model divides the GW signal into three stages: inspiral, intermediate, and ringdown. In this work, we employ the PhenomD model to represent the inspiral and intermediate waveforms. To model the ringdown waveform, we introduce the $\Phi$ model, which we explain in detail later in the paper.

\subsection{inspiral}
The phase ansatz in the inspiral stage is given by

\begin{equation}
\begin{aligned}
\phi_{\mathrm{Ins}}= & \phi_{\mathrm{TF} 2}(M f ; \Xi) \\
& +\frac{1}{\eta}\left(\sigma_{0}+\sigma_{1} f+\frac{3}{4} \sigma_{2} f^{4 / 3}+\frac{3}{5} \sigma_{3} f^{5 / 3}+\frac{1}{2} \sigma_{4} f^{2}\right)\\
&+\phi_{\mathrm{KRZ}}
\end{aligned}
\end{equation}

where $\eta=m_1 m_2/M^2$, $M = m_1+m_2$, the $\phi_{\mathrm{TF} 2}$ is the full TaylorF2 phase:
\begin{equation}\label{Ins_equ_D}
\begin{aligned}
\phi_{\mathrm{TF} 2}= & 2 \pi f t_{c}-\varphi_{c}-\pi / 4 \\
& +\frac{3}{128 \eta}(\pi f M)^{-5 / 3} \sum_{i=0}^{7} \varphi_{i}(\Xi)(\pi f M)^{i / 3}
\end{aligned}
\end{equation}
The constants $\sigma_{i}$ (where $i = 0, 1, 2, 3, 4$) represent the correlation between the mass and spin of the system. Meanwhile, the phase deformation arising from the general parameterized black hole is denoted by $\phi_{\mathrm{KRZ}}$. Varying the values of $\delta_1$, $\delta_2$, $\delta_4$, and $\delta_6$ will result in different phases. $\varphi_{i}(\Xi)$ are the PN expansion coefficients that are related to the intrinsic binary parameters. The detailed information of $\sigma_{i}$ and $\varphi_{i}(\Xi)$ can be found in Appendix B of this article\cite{phenomD}. It should be mentioned here that the inspiral phase relies on the PN results which only have finite correction orders, so it may miss the strong-field effects at higher PN orders that appear in some modified theories.

\subsection{Intermediate}
The intermediate stage is after the inspiral stage, and its phase is given by\cite{phenomD}
\begin{equation}\label{Int_equ_D}
\phi_{\mathrm{Int}}=\frac{1}{\eta}\left(\beta_0+\beta_1 f+\beta_2 \log (f)-\frac{\beta_3}{3} f^{-3}\right)
\end{equation}
$\beta_{i}(i=0, 1, 2, 3)$ is the constants related to the mass and spin of the system. The detailed information of $\beta_{i}$ can still be found in Appendix B of this article\cite{phenomD}. Because the duration of the intermediate is indeed short and the accuracy of Eq.~(\ref{Int_equ_D}) is good enough compared with the numerical relativity (NR) waveforms\cite{phenomD}, we directly use this phase to construct our waveform model and ignore the non-GR effect.

\section{The ringdown waveform in KRZ Black Holes}\label{Ringdown}

The forthcoming section aims to capture the ringdown signals from the waveforms generated using the PSI ($\Psi$:photon sphere + inspiral) model \cite{Psi_22}. This model operates in the proximity of the photon sphere and is based on the BOB waveform model first presented in Ref.~\cite{McWilliams_19}. The BOB model is an analytic phenomenological model for the late inspiral, merger, and ringdown signal of binary black hole (BBH). It takes into account the motion of photons without considering any additional phenomenological degrees of freedom. This paper aims to derive the complete waveforms by associating our $\Psi$ waveforms with the inspiral waveforms. We will then scrutinize the complete waveforms with diverse spins and mass ratios and apply them to the parametrized black holes(i.e., KRZ metric).

To utilize the $\Psi$ waveforms, it is essential to derive the parameters of QNMs, i.e., $\omega_R$ and $\omega_I$. The real part of QNMs, $\omega_R$, can be decomposed into two directional components, namely, $\theta$ and $\phi$:
\begin{equation}
\omega_R=L\Omega_{\theta}(m/L)+m\Omega_{\mathrm{prec}}(m/L)
\end{equation}
where $\Omega_{\theta}$ indicates the frequency of polar motion, which is the rate at which the photon oscillates above and below the equatorial plane. The oscillation period can be calculated using the formula, $T_{\theta}=2\pi/\Omega_{\theta}$.

In addition to polar motion, the particle also undergoes a periodic motion in the azimuthal ($\phi$) direction with respect to the oscillation period, $T_{\theta}$, and the magnitude, $\Delta \phi$. The deviation between $\Delta \phi$ and $\pm 2\pi$ is commonly known as the "precession angle":
\begin{equation}
\Delta \phi_{\mathrm{prec} }=\Delta \phi -\begin{cases}
-4\pi\quad(\mathrm{corotating \; orbit} ) \\
+4\pi\quad(\mathrm{couterrotating \; orbit})
\end{cases}
\end{equation}
\begin{equation}
\Omega_{\mathrm{prec}}=\Delta\phi_{\mathrm{prec} }/T_{\theta}
\end{equation}
\begin{equation}
L=l+1/2.
\end{equation}

\begin{figure*}
\centering
\includegraphics[width=0.45 \textwidth]{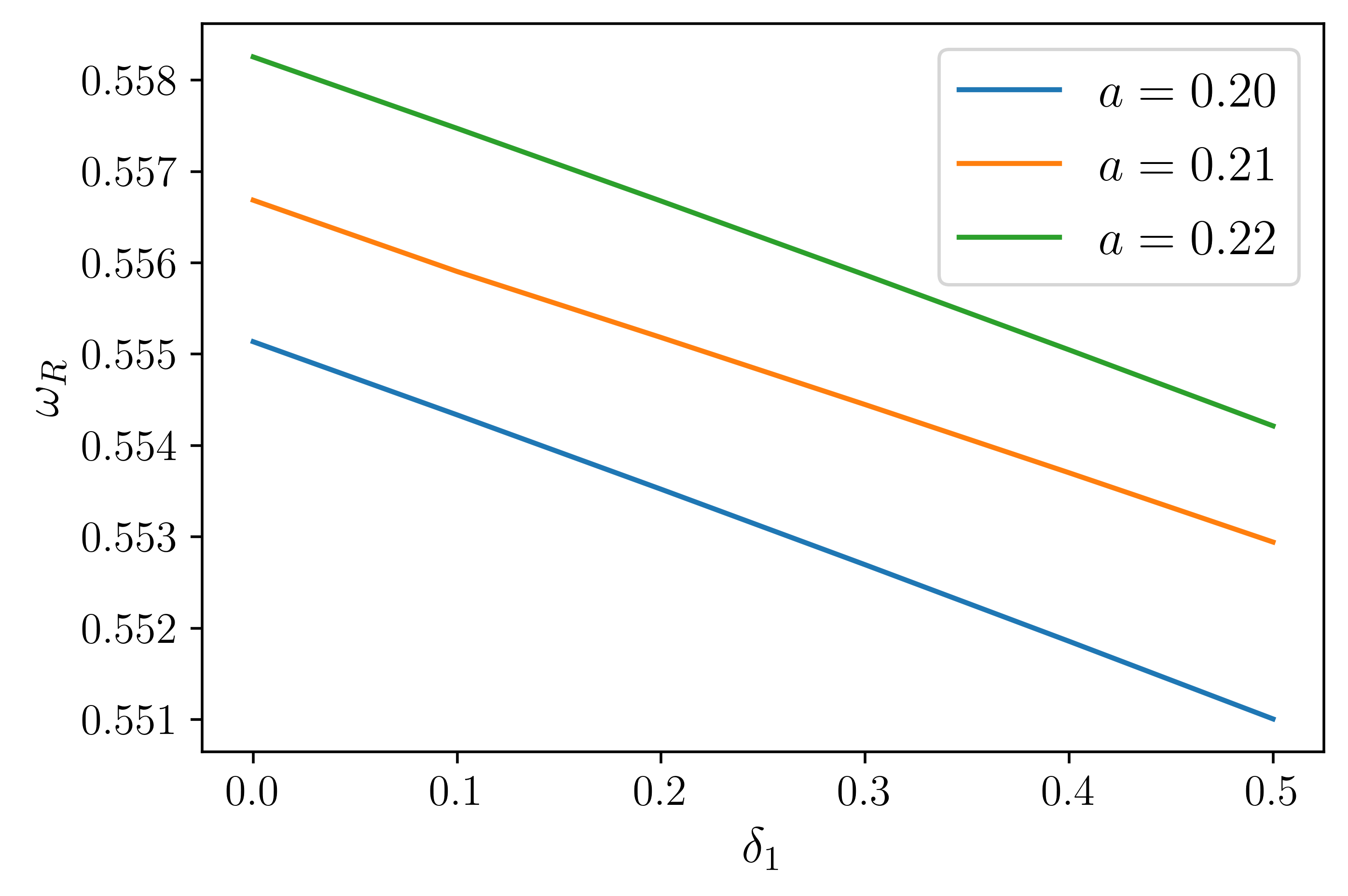}
\includegraphics[width=0.45 \textwidth]{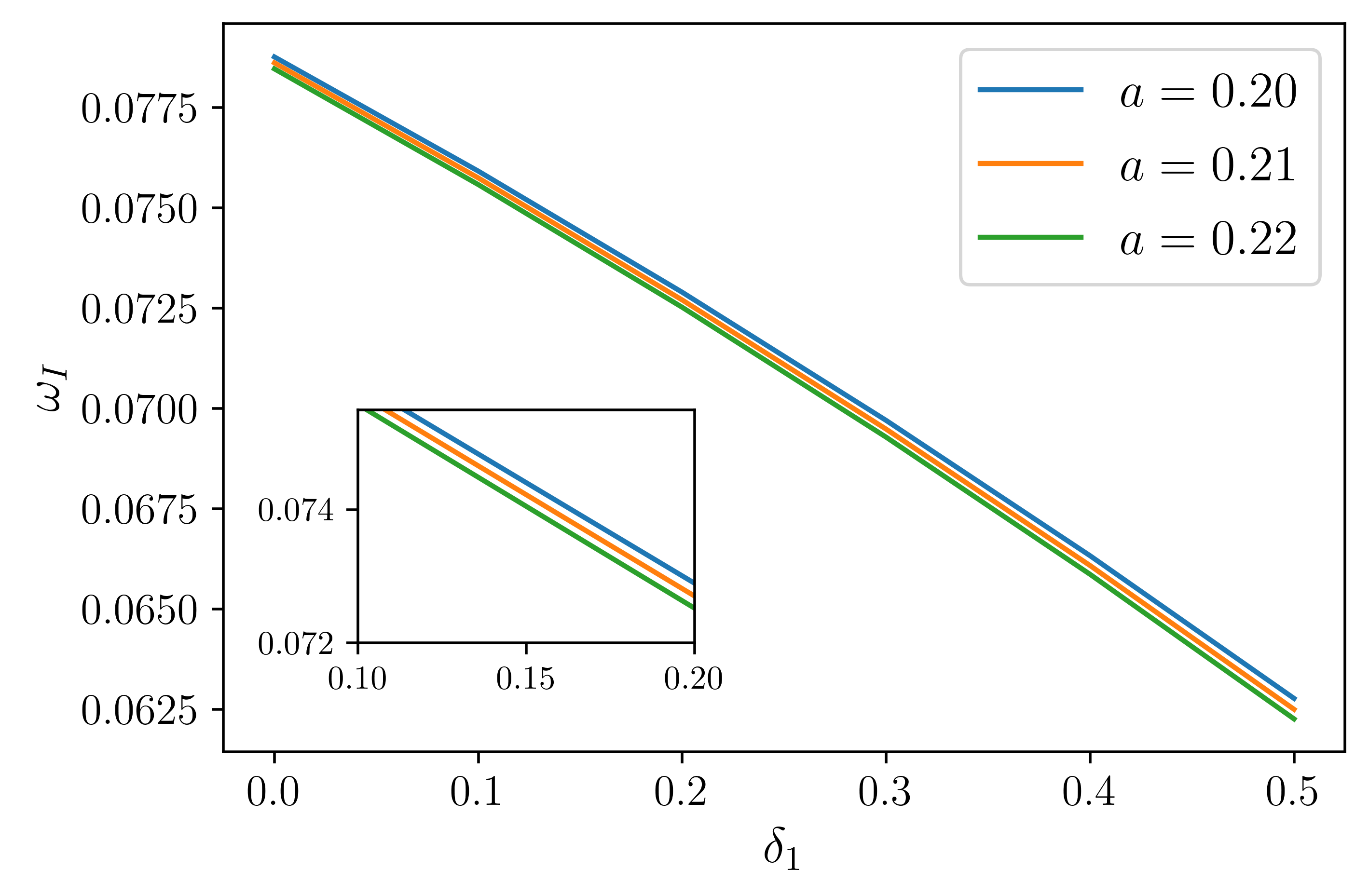}
\includegraphics[width=0.45 \textwidth]{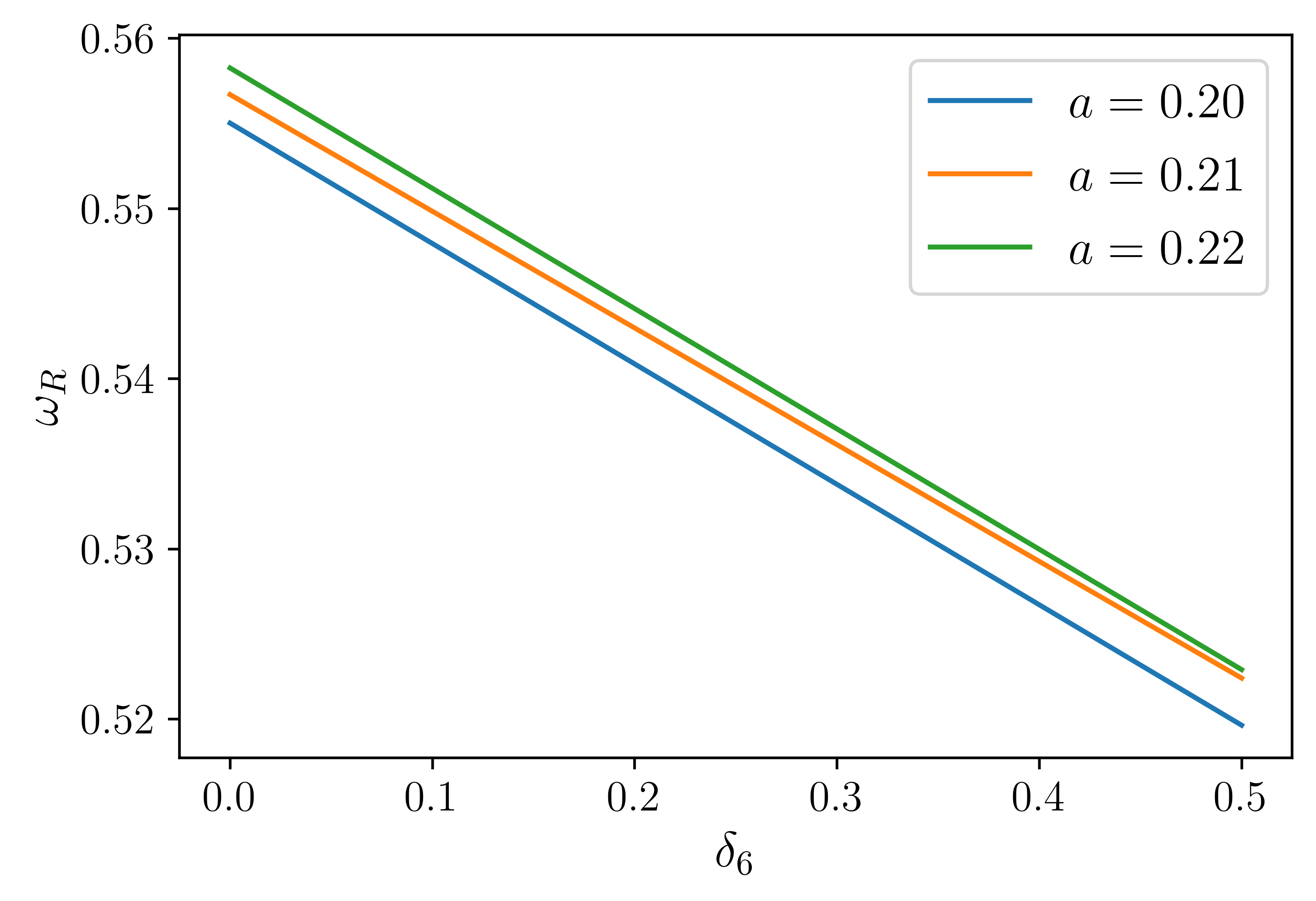}
\includegraphics[width=0.45 \textwidth]{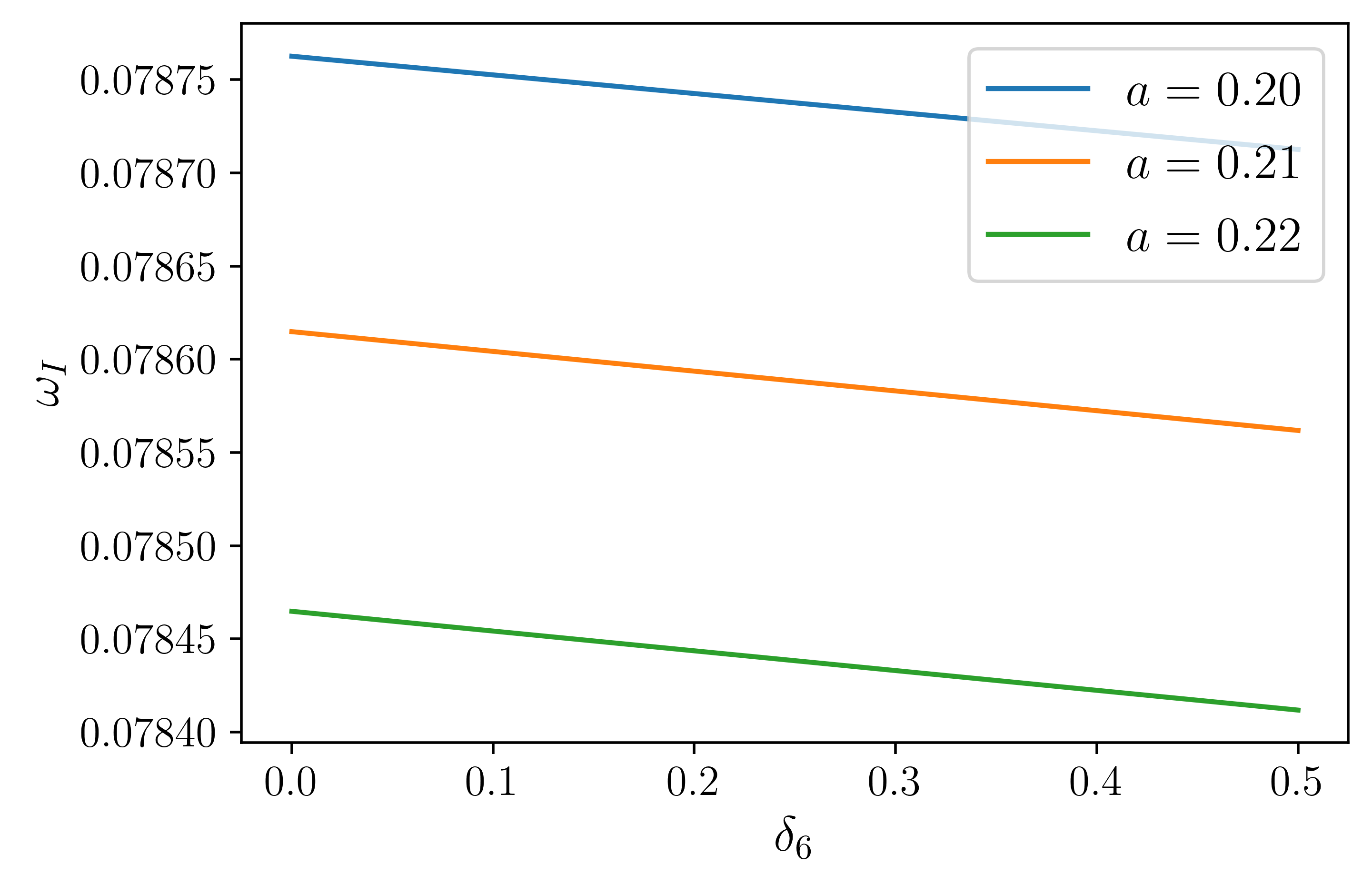}
\caption{The relationship between the $\omega_{R(I)}$ and $\delta_{1(6)}$ with different spin $a$($a=0.20, 0.21, 0.22$). The left panel shows the relationship between the $\omega_R$ and $\delta_{1(6)}$, and the right panel shows the relationship between the $\omega_I$ and $\delta_{1(6)}$. When we study one of the parameters $\delta_i$ another parameter is equal to zero.}\label{OmegaRI_delta16}
\end{figure*}

The values of $l$ and $m$ could be determined through the conditions:$V^{r}\left(r, \omega_{R}\right)=\left.\frac{\partial V^{r}}{\partial r}\right|_{\left(r, \omega_{R}\right)}=0$, and $V^r$ is the potential in the radial Teukolsky equation. The imaginary component of QNMs, $\omega_I$, is directly linked to the Lyapunov exponents, which determine the rate at which a circular null geodesic expands its cross-sectional area under infinitesimal radial perturbations. The detailed calculation of $\omega_I$ can be found in Ref.~\cite{Yang_12}. We can derive $\omega_R$ and $\omega_I$ using the photon motion in three dimensions (3D). Figure~\ref{OmegaRI_delta16} illustrates the correlation between the real and imaginary part of the QNM frequency $\omega_R$, $\omega_I$ and between the deformation parameters $\delta_1$, $\delta_6$. Varying $\delta_6$ has a more significant effect on $\omega_R$. In the figure at the bottom left, we note that the green and orange lines almost overlap, beyond a spin value, but since $\delta_6$ cannot exceed $0.5$, this will not happen. This is caused by the quadrupole moment deviation $\Delta Q$ becoming excessively large. On the right panel, it is evident that the parameter $\delta_1$ significantly impacts the value of $\omega_I$.

\begin{figure*}
\centering
\includegraphics[width=0.45 \textwidth]{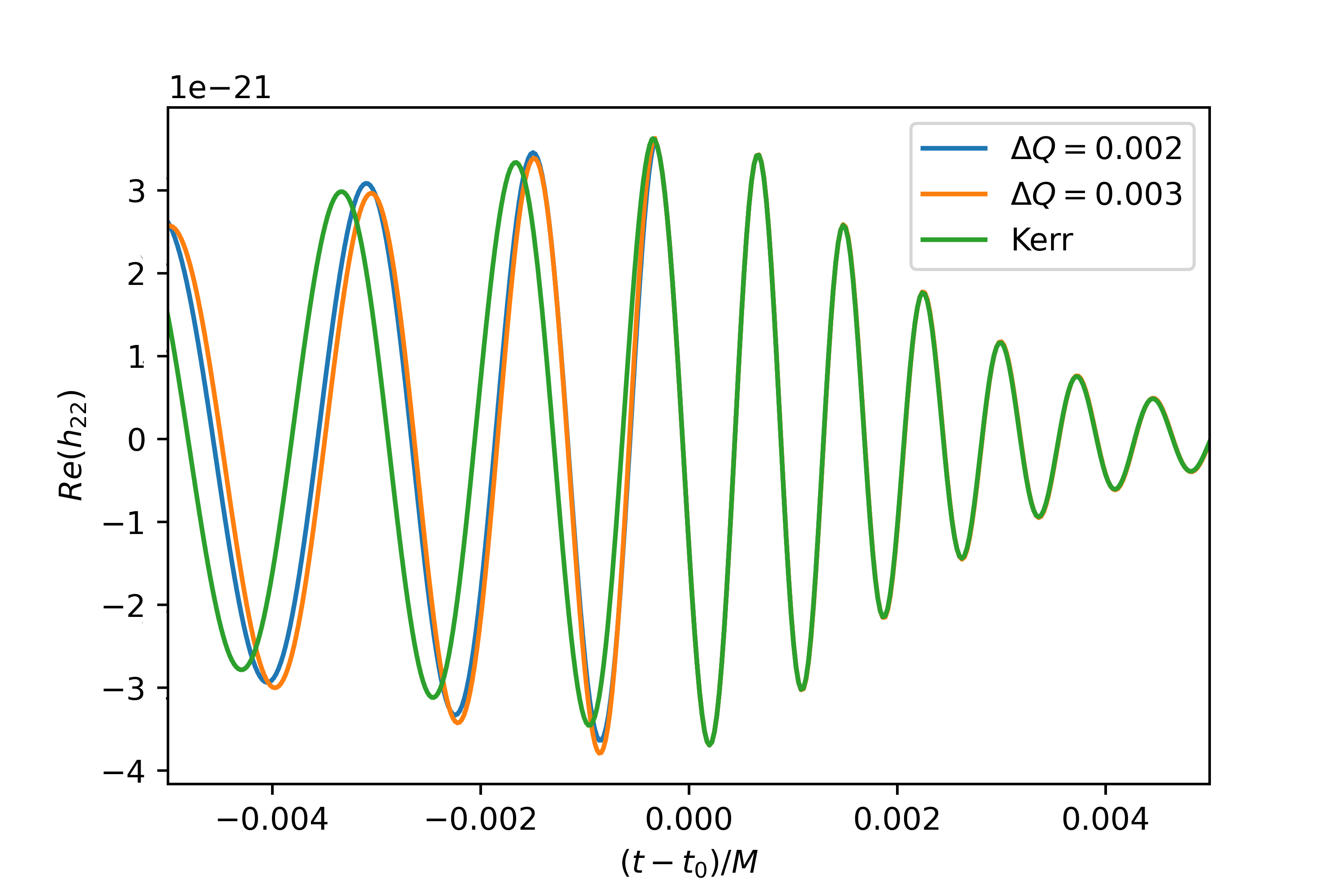}
\includegraphics[width=0.45 \textwidth]{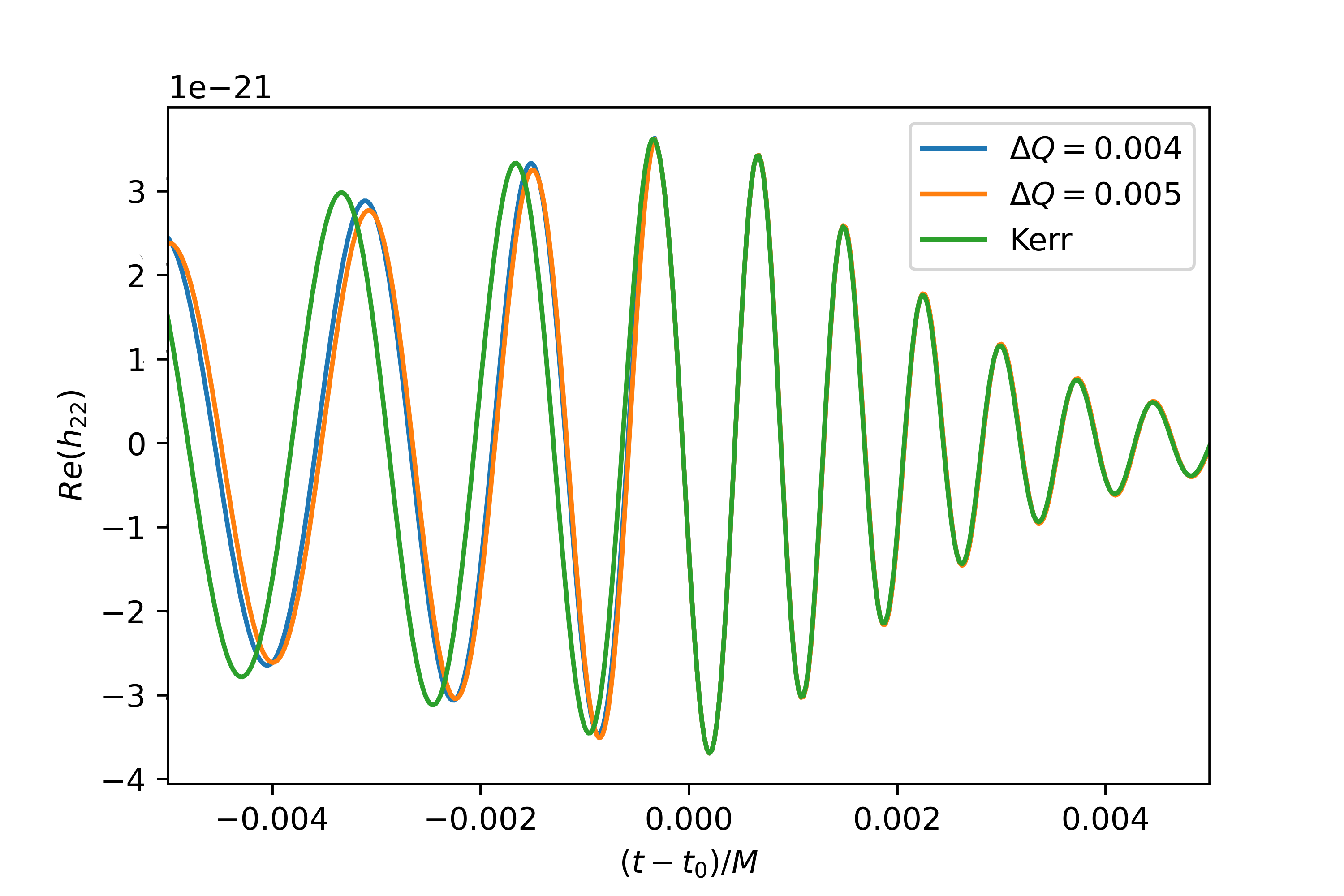}
\includegraphics[width=0.45 \textwidth]{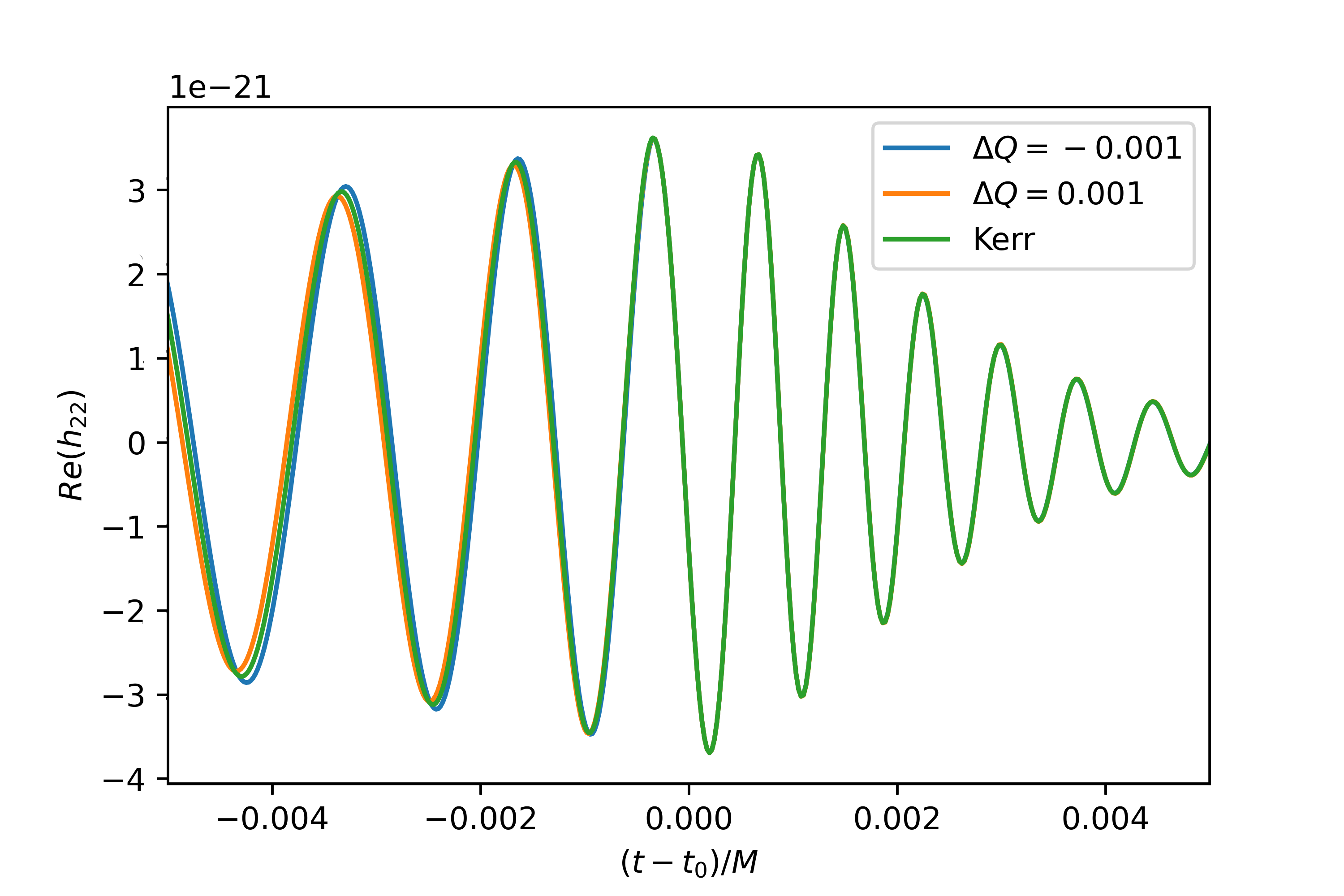}
\includegraphics[width=0.45 \textwidth]{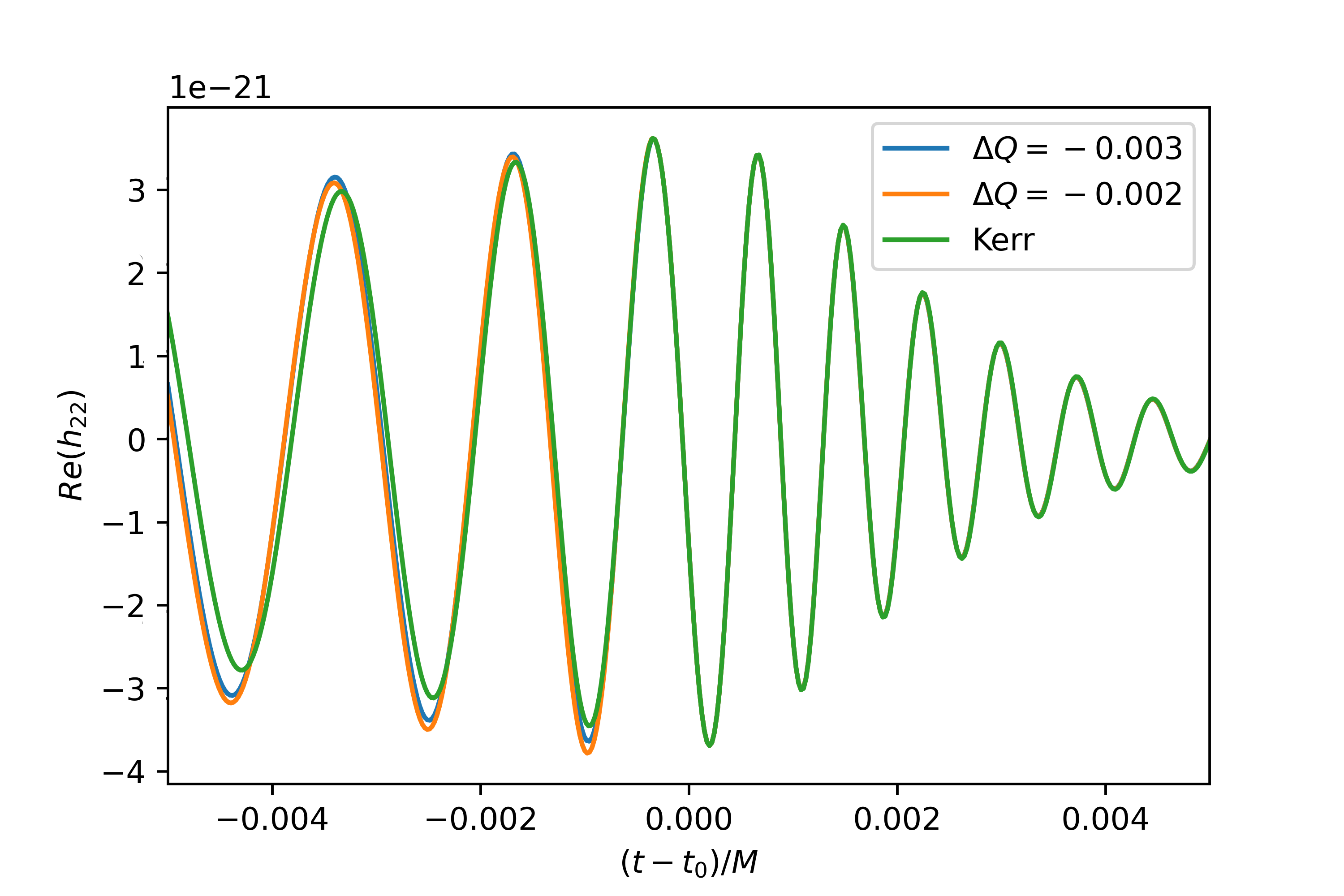}
\caption{The full waveform of the bumpy black hole with spin $\chi_1=\chi_2=0.85$ and mass ratio $1:1$. Each panel shows the different waveform with different quadrupole moment deviations $\Delta Q$. For better comparison, we chose different values that had both positive and negative values of quadrupole moment deviations $\Delta Q$.}\label{full_a085}
\end{figure*}

\begin{figure*}
\centering
\includegraphics[width=1.00 \textwidth]{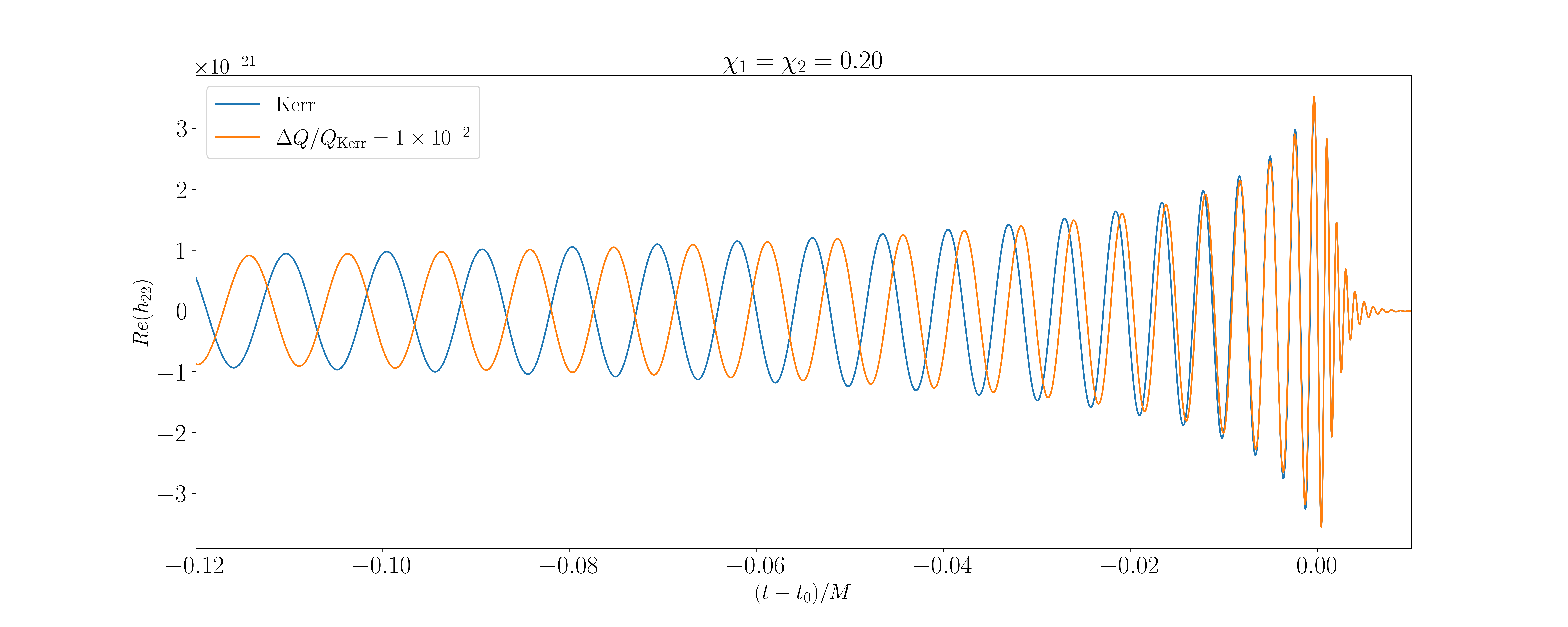}
\includegraphics[width=1.00 \textwidth]{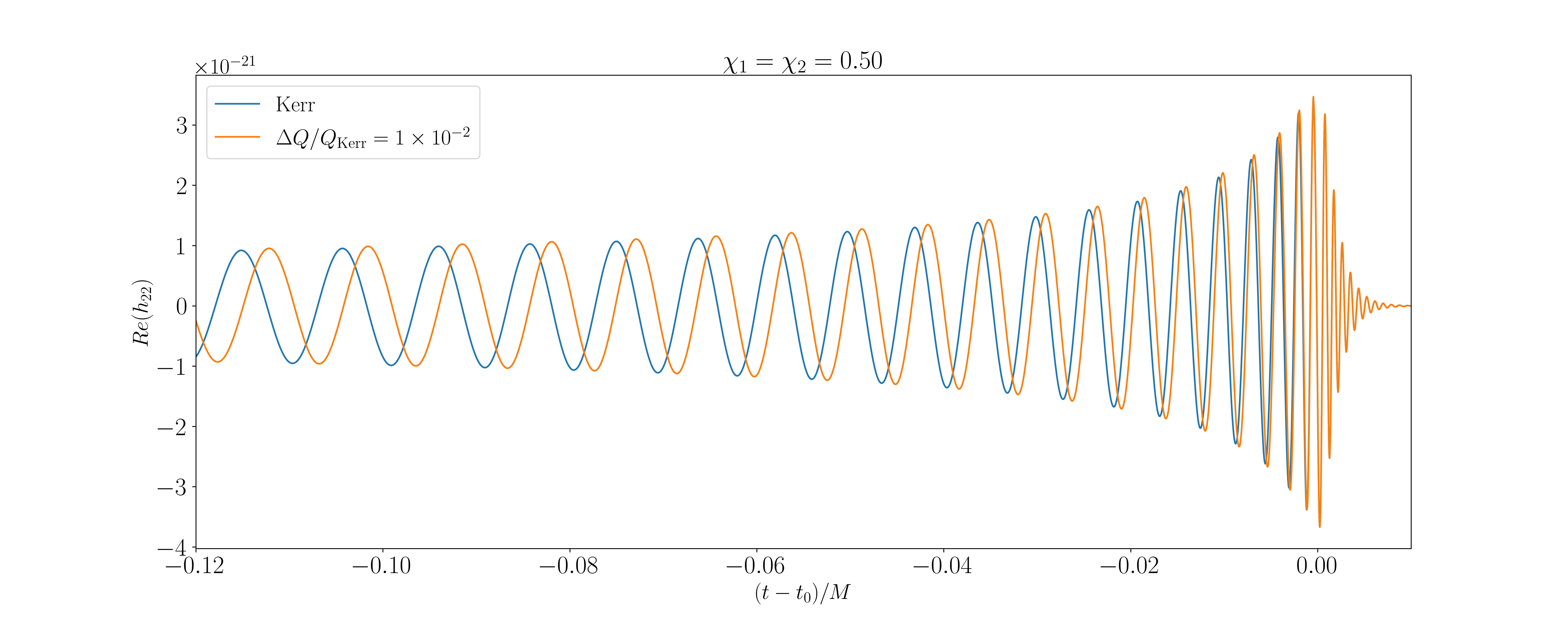}
\includegraphics[width=1.00 \textwidth]{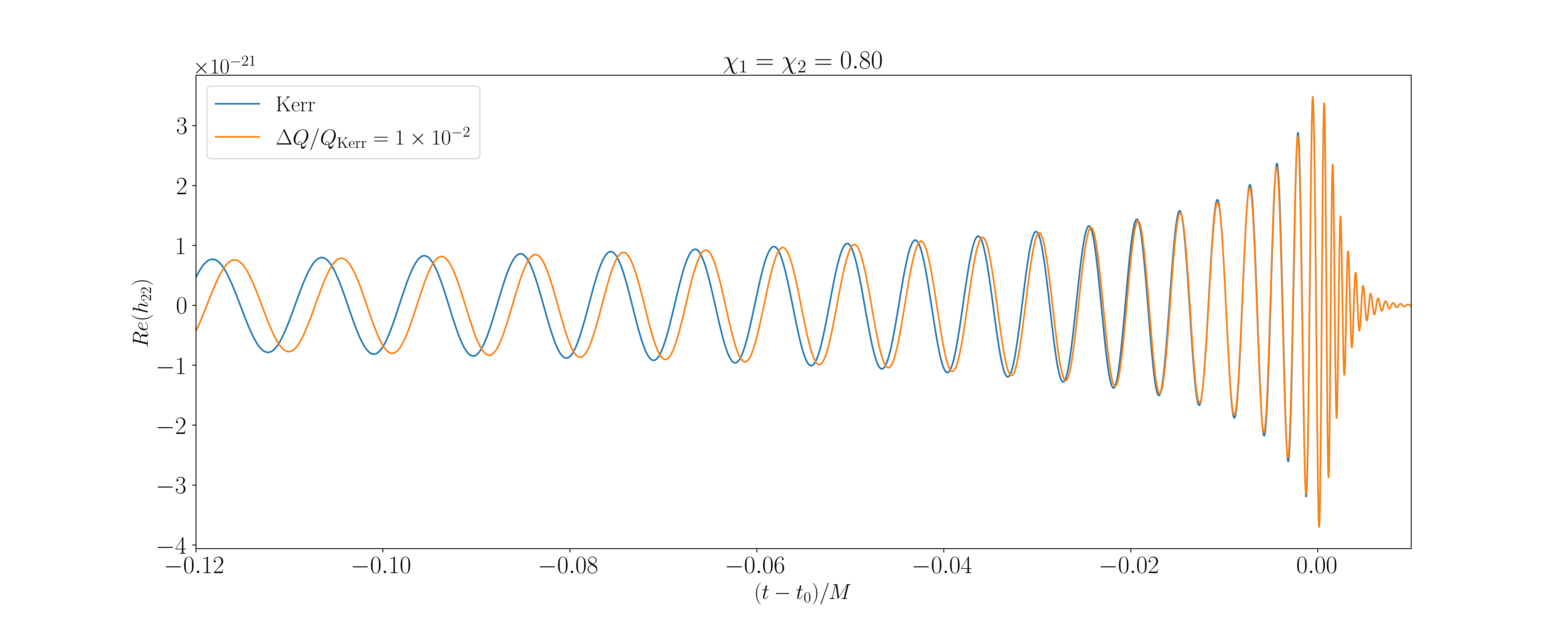}
\caption{The full waveform of the bumpy black hole with a different spin($\chi_1=\chi_2=0.20, 0.50$ and $0.80$) and with the same mass ratio $1:1$. Each panel shows the waveform with the same relative quadrupole moments $\Delta Q/ Q_{\mathrm{Kerr}}= 1 \times 10^{-2}$ and the same spin Kerr waveform.}\label{full_GW}
\end{figure*}

\begin{figure*}
\centering
\includegraphics[width=0.45 \textwidth]{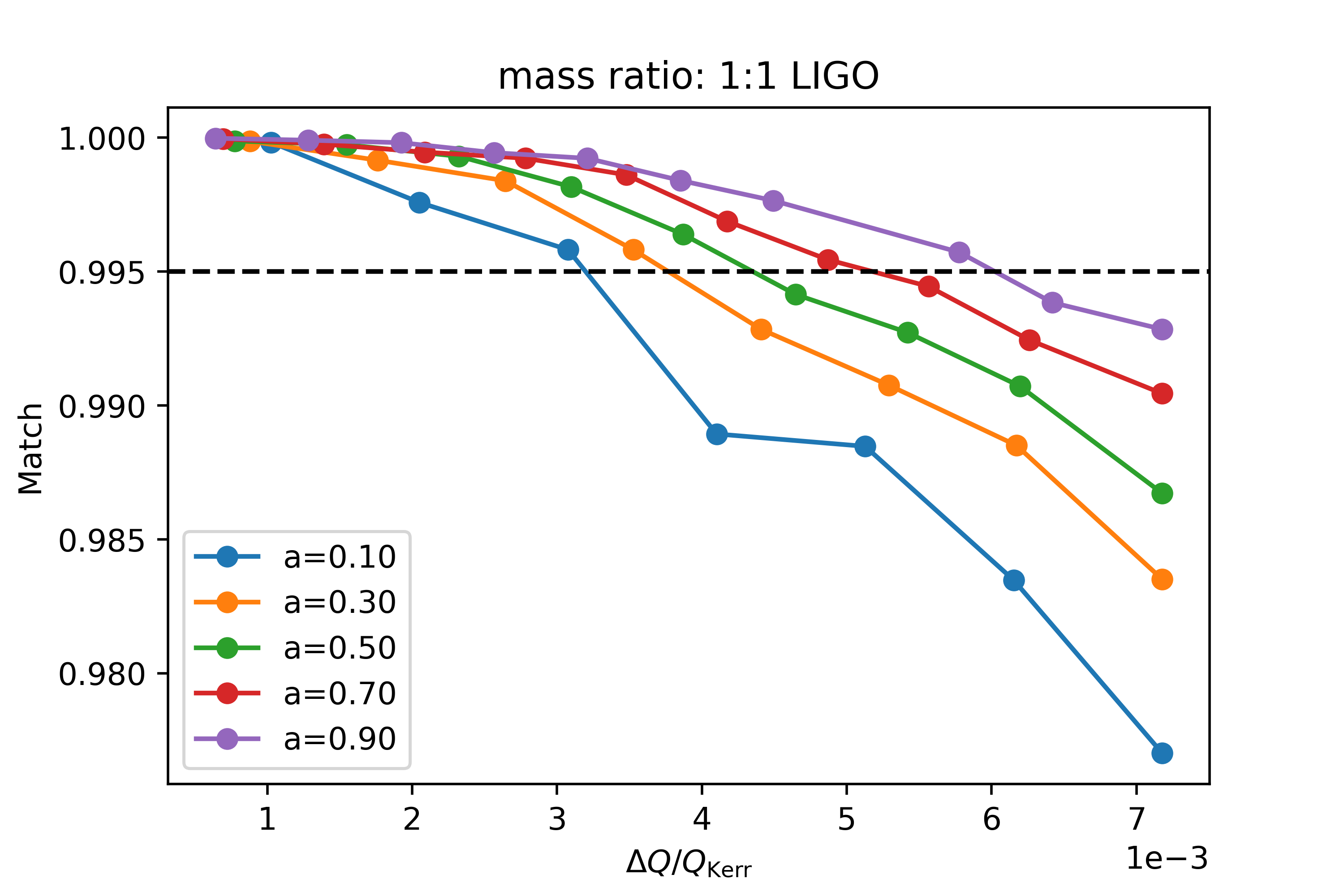}
\includegraphics[width=0.45 \textwidth]{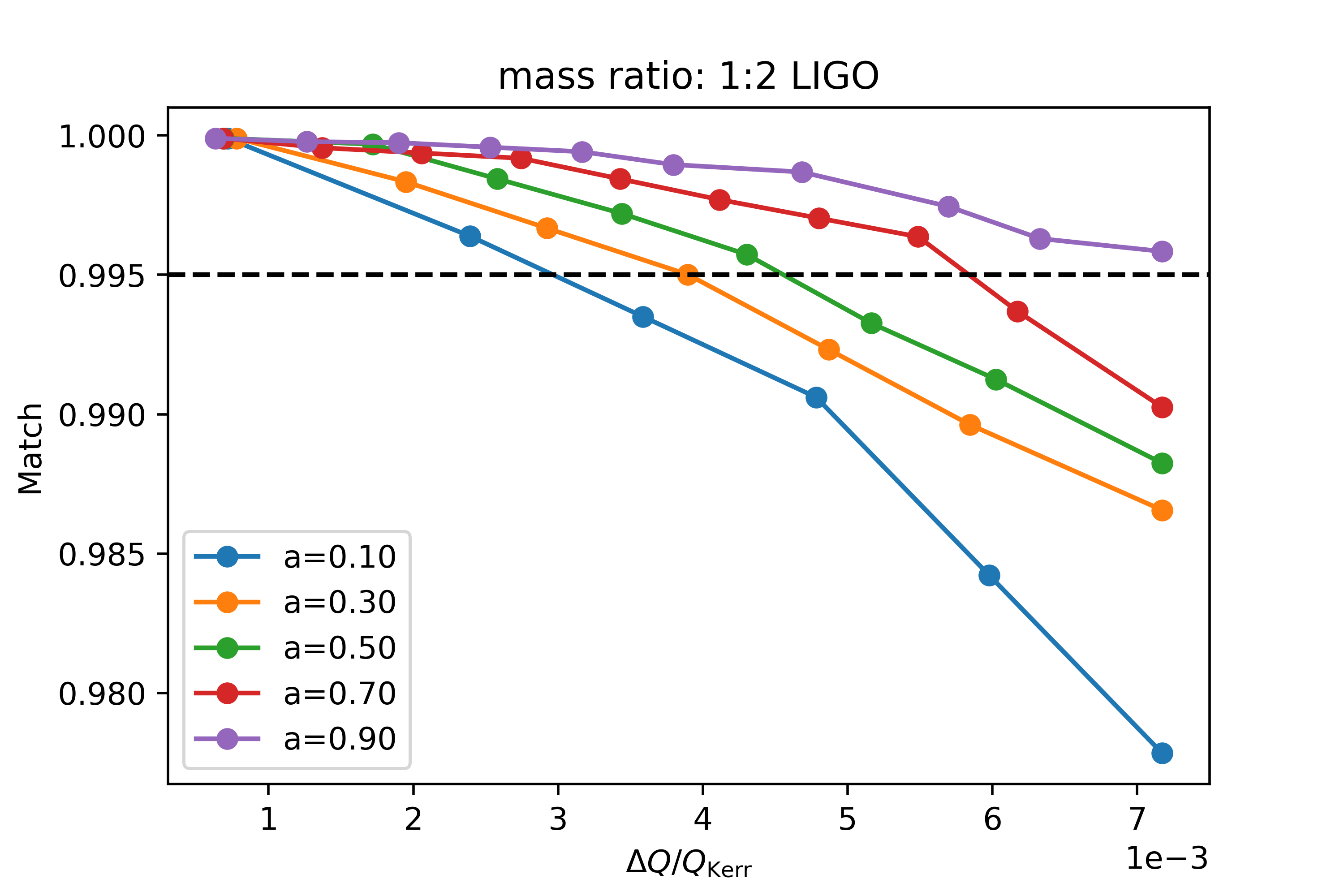}
\includegraphics[width=0.45 \textwidth]{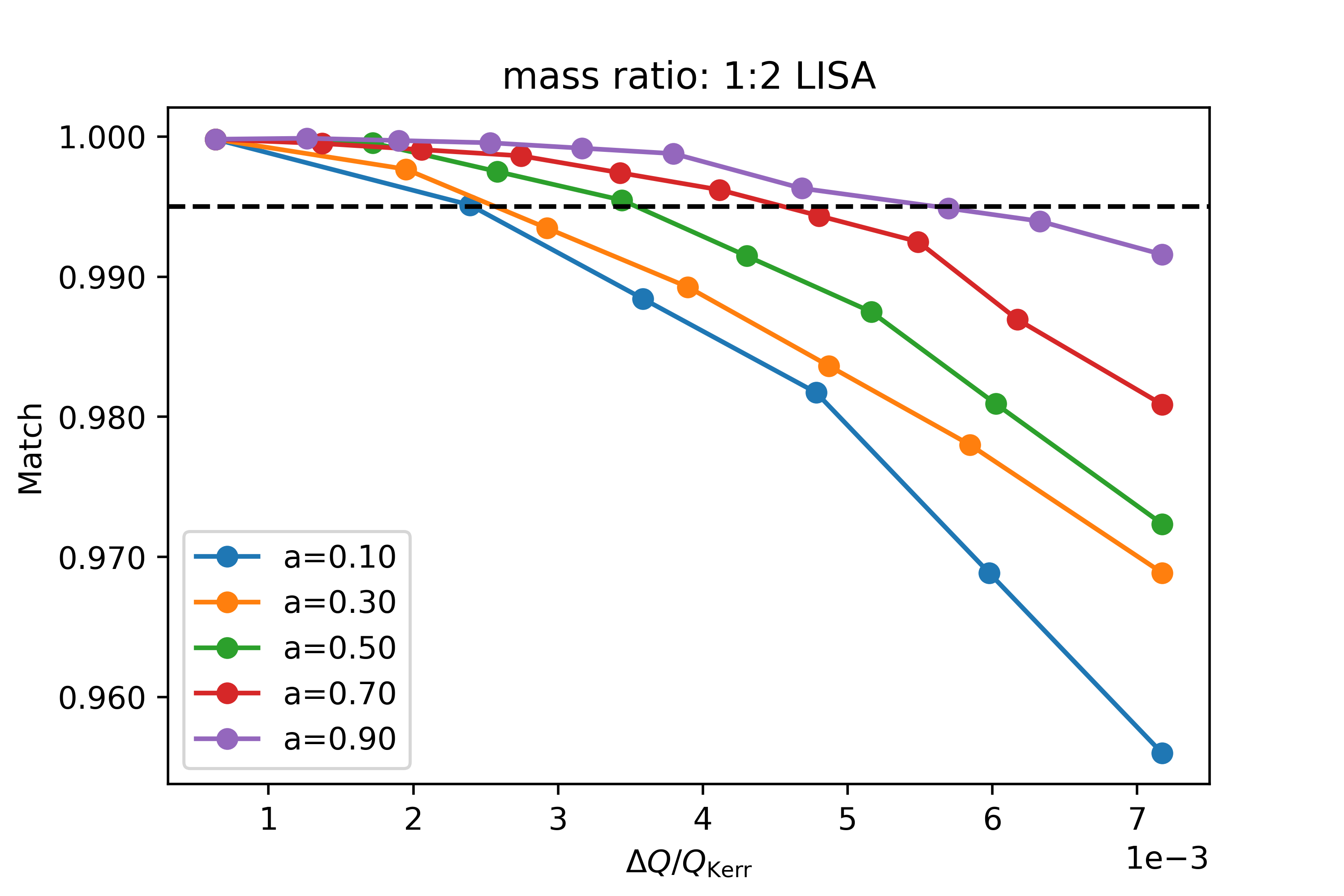}
\includegraphics[width=0.45 \textwidth]{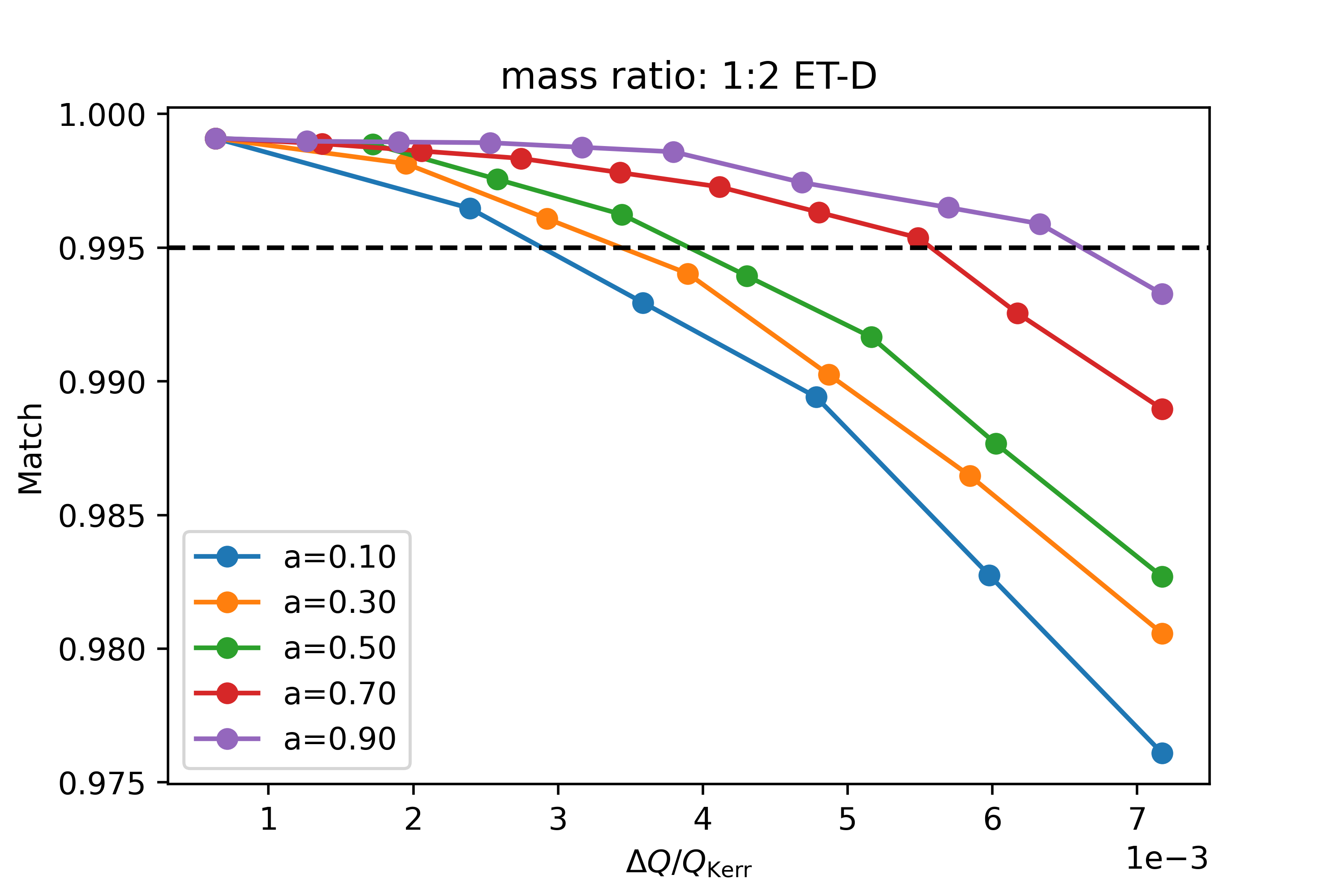}
\caption{The match between the Kerr waveform and the bumpy black hole waveform with different relative quadrupole moment $\Delta Q/Q_{\mathrm{Kerr}}$. In each panel, we plot the cases with different spins $a=0.10, 0.30, 0.50, 0.70$, and $0.90$. The left panel shows the case of mass ratio $1:1$, and the right panel shows the mass ratio $1:2$. The dashed black line represents the match values equal to 0.995.}\label{Match}
\end{figure*}

The amplitude of the gravitational wave can be expressed as follows, based on \cite{McWilliams_19} \cite{BOB_b}
\begin{equation}
{{\left| {{h}_{lm}} \right|}^{2}}\sim \frac{d}{dt}\left( {{\Omega }_{lm}}^{2} \right)\,,
\end{equation}
where $\Omega_{lm}$ is the orbital frequency, through this equation, and we can get the equation of the GW waveform:

\begin{equation}\label{RD_equ_D}
{h}_{22}=X \operatorname{sech}\left[\gamma\left(t-t_{p}\right)\right] e^{-i \tilde{\Phi}_{22}(t)}\,.
\end{equation}
The equation includes the following variables: $X$ is a constant related to the amplitude of the waveform, $\gamma$ is the Lyapunov exponent characterizing the rate of divergence of nearby null geodesics, $t_p$ is the time at maximum amplitude of the waveform and $\Phi_{22}(t)$ is the phase. 

We can also derive the phase equation:
\begin{equation}
\begin{aligned}
\tilde{\Phi}_{22}=& \int_{0}^{t} \Omega d t^{\prime}=\arctan _{+}+\operatorname{arctanh}_{+} \\
&-\arctan _{-}-\operatorname{arctanh}_{-}-\phi_0,
\end{aligned}
\end{equation}

where
\begin{equation}
\left\{\begin{array}{c}
\arctan _{\pm} \equiv \kappa_{\pm} \tau\left[\arctan \left(\frac{\Omega}{\kappa_{\pm}}\right)-\arctan \left(\frac{\Omega_{0}}{\kappa_{\pm}}\right)\right]\,, \\
\arctan \mathrm{h}_{\pm} \equiv \kappa_{\pm} \tau\left[\arctan h\left(\frac{\Omega}{\kappa_{\pm}}\right)-\arctan h\left(\frac{\Omega_{0}}{\kappa_{\pm}}\right)\right]
\end{array}\right.
\end{equation}
\begin{equation}
\kappa_{\pm} \equiv\left\{\Omega_{0}^{4} \pm k\left[1 \mp \tanh \left(\frac{t_{0}-t_{p}}{\tau}\right)\right]\right\}^{1 / 4}\,,
\end{equation}

\begin{equation}\label{attention1}
\Omega=\left\{\Omega_{0}^{4}+k\left[\tanh \left(\frac{t-t_{p}}{\tau}\right)-\tanh \left(\frac{t_{0}-t_{p}}{\tau}\right)\right]\right\}^{1 / 4}\,,
\end{equation}

\begin{equation}\label{attention2}
k=\left(\frac{\Omega_{\mathrm{QNM}}^{4}-\Omega_{0}^{4}}{1-\tanh \left[\left(t_{0}-t_{p}\right) / \tau\right]}\right)\,,
\end{equation}\\
where $\tau=\gamma^{-1}$, ${{\Omega }_{\operatorname{QNM}}}$=$\omega_{\operatorname{QNM}}$/m($\Omega_{\operatorname{QNM}}$ is just $\omega_R$), and $\phi_0$, $\Omega_0$, $t_0$ are the constants that can be freely chosen.

We need to focus on Eqs.~(\ref{attention1}) and (\ref{attention2}) mentioned above. The inclusion of terms with an even power in these equations imposes an extra constraint on $\Omega_0$. Our objective is to determine the minimum value of $\Omega_0$, which can be achieved by equating the expression inside Eq.~(\ref{attention1}) to zero. This yields the following function:
\begin{equation}\label{attention3}
\Omega_{0}^{4}=k\left[-\tanh \left(\frac{t-t_{p}}{\tau}\right)+\tanh \left(\frac{t_{0}-t_{p}}{\tau}\right)\right]\,.
\end{equation}
Substituting Eq.~(\ref{attention2}) into Eq.~(\ref{attention3}), we can get the solution of Eq.~(\ref{attention3})(we only consider the positive solution):
\begin{equation}\label{Omega_min}
{{\Omega }_{0}}^{4}=\frac{{{\Omega }_{\text{QNM}}}^{4}(\tanh [\frac{t-{{t}_{p}}}{\tau }]-\tanh [\frac{{{t}_{0}}-{{t}_{p}}}{\tau }])}{(-1+\tanh [\frac{{{t}_{0}}-{{t}_{p}}}{\tau }])(1-\frac{\tanh [\frac{t-{{t}_{p}}}{\tau }]}{1-\tanh [\frac{{{t}_{0}}-{{t}_{p}}}{\tau }]}+\frac{\tanh [\frac{{{t}_{0}}-{{t}_{p}}}{\tau }]}{1-\tanh [\frac{{{t}_{0}}-{{t}_{p}}}{\tau }]})}\,.
\end{equation}\\

With Eq.~(\ref{Omega_min}), we get the minimum value of $\Omega_0$. For convenience, we choose $t$ equal to $t_p$, so Eq.~(\ref{Omega_min}) can be simplified to this form:
\begin{equation}\label{Omega_min_simplify}
{{\Omega }_{0}}^{4}={{\Omega }_{\text{QNM}}}^{4}(\tanh [\frac{{{t}_{0}}-{{t}_{p}}}{\tau }])\,.
\end{equation}\\
Thus, we obtain the minimum value of $\Omega_0$ is $\Omega_{\operatorname{QNM}}$(i.e. the region of $\Omega_0$ is $\Omega_0 \textgreater \Omega_{\operatorname{QNM}}$).

To finalize the analysis, it is essential to establish a connection between the waveforms originating from the photon sphere and the inspiral waveforms. This can be established at any point between the innermost stable circular orbit (ISCO) and the light ring (LR). In this study, we have selected the peak of the waveform as our matching point. This selection allows us to derive the optimal values of $\phi_0$, $\Omega_0$, and $t_0$. Then with Eq.~(\ref{Ins_equ_D}), (\ref{Int_equ_D}), and (\ref{RD_equ_D}), we can get the full waveform and refer to it as PSI-FD($\Psi_{\mathrm{FD}}$).

In Fig.~\ref{full_a085}, we show the full waveform for the spin $\chi_1=\chi_2=0.85$, for mass ratio $1:1$, and with different quadrupole moment deviation $\Delta Q$. The waveform from Kerr black holes is plotted in each panel for comparison purposes. Both positive and negative values of deviations $\Delta Q$ were selected to ensure a more thorough comparison. The waveforms suggest that the quadrupole moment deviation denoted as $\Delta Q$ has a minor impact on the ringdown part. However, it significantly influences the inspiral part. The analysis of Figs.~\ref{DeltaQ_delta_Bumpy} and \ref{OmegaRI_delta16} can explain that for the ringdown part, a major influence is from the values of $\omega_R$ and $\omega_I$. Further analysis of these figures identifies that the $\omega_{R(I)}$ only undergoes minor changes with a $\Delta Q$ magnitude variation of $0.001$. The full waveforms of the same relative quadrupole moment $\Delta Q/Q_{\mathrm{Kerr}}$ in different spins are shown in Fig.~\ref{full_GW}. Our analysis indicates that at a small spin, the quadrupole moment deviation $\Delta Q$ has a considerably greater impact on the overall waveform. Furthermore, to ensure a more intuitive comparison, we employ the parameter overlap. The definition of the overlap is as follows:

\begin{equation}
\mathrm{F}= \left[\frac{\left\langle h_{1} \mid h_{2}\right\rangle}{\sqrt{\left\langle h_{1} \mid h_{1}\right\rangle\left\langle h_{2} \mid h_{2}\right\rangle}}\right],
\end{equation}

\begin{equation}
\left\langle h_{1}, h_{2}\right\rangle=4 \operatorname{Re} \int_{f_{\min }}^{f_{\max }} \frac{\tilde{h}_{1}(f) \tilde{h}_{2}^{*}(f)}{S_{n}(f)} d f,
\end{equation}

where $h_1$ is the waveform derived from the PSI model, $h_2$ is the compared waveform(e.g., SEOBNRv4, SXS), and $S_n(f)$ is the power spectral density of the detector noise; in this work we use the aLIGO's sensitivity curve\cite{Harry_2010}. And the definition of the match is
\begin{equation}
\mathrm{FF}=\max \left[\frac{\left\langle h_{1} \mid h_{2}\right\rangle}{\sqrt{\left\langle h_{1} \mid h_{1}\right\rangle\left\langle h_{2} \mid h_{2}\right\rangle}}\right],
\end{equation}
and the mismatch of two waveforms is defined as $1-\mathrm{FF}$. Figure~\ref{Match} displays the plot of the match between the Kerr waveforms and the bumpy black hole with various relative quadrupole moments $\Delta Q/Q_{\mathrm{Kerr}}$. The plots for each panel represent different spin values, specifically $a=0.10$, $0.30$, $0.50$, $0.70$, and $0.90$. The dashed black line represents the match values equal to 0.995 which corresponds to the SNR approximately equal to $40$, and this value is from the definition $F=1-\frac{D}{2 \mathrm{SNR}_{min}^{2}}$ where $F$ is the match between the signals, and $D$ is the number of intrinsic parameters of the model(we choose $D=14$ in this work). The left panel shows the case of mass ratio $1:1$, and the right panel shows the mass ratio $1:2$. The match exhibits a smaller value at lower spin configurations, which can be explained using the definition of the quadrupole moment. According to Sec.~\ref{KRZ}, the definition of the quadrupole moment $Q$ can be written as
\begin{eqnarray}
Q & = & -\frac{2_{20}r_{0}^3+Ma^2}{3},
\end{eqnarray}
where $r_0$ is the horizon radius. Then we can derive this expression:
\begin{eqnarray}
\frac{Q_{\text {Kerr }}-Q}{Q_{\mathrm {Kerr }}}  =  \frac{\Delta Q}{Q_{\mathrm {Kerr }}} =\frac{2}{3}-a_{20} \frac{r_0^{3}}{M a^{2}}
\end{eqnarray}
where the parameter $a_{20}$ is directly proportional to the deformation parameters $\delta_i$. It is evident that as the value of $\delta_i$ increases, the deviation from the Kerr case becomes more pronounced, leading to a decrease in the match values. Based on this, we can rewrite the above equations as follows:

\begin{eqnarray}
\mathrm{Match}\propto  (\frac{2}{3}-\frac{\Delta Q}{Q_{\mathrm{Kerr} }})\frac{Ma^2}{r_0^3}
\end{eqnarray}

From this equation, we see that when the value of $\Delta Q/Q_{\mathrm{Kerr}}$ is fixed, a smaller value of $a$ results in a smaller match(the difference of $a^2/r_{0}^3$ is really small for the different spins).

We also study the overlap between the Kerr waveform and $\Psi_{\mathrm{FD}}$ for the next-generation detectors such as Laser Interferometer Space Antenna(LISA) and Einstein Telescope(ET) as shown in Fig.~\ref{Match}. The bottom panel shows the overlap with different relative quadrupole moments $\Delta Q/Q_{\mathrm{Kerr}}$ for LISA\cite{LISA} and ET-D sensitivity\cite{ET_D} of the same mass ratio $1:2$. They have the same trend as shown in LIGO where the overlap exhibits a smaller value at lower spin configurations. Compared with the top panel, we can see the overlap has a smaller value for the LISA and ET, and this means the next generation will have a better detection of such a gravitational waveform.

\section{Conclusion}\label{Conclusion}
In this article, we investigate a general parametrization of axisymmetric black holes, using the KRZ metric. Previous studies have controversially defined the equatorial radius of the event horizon, $r_0$, based on the Kerr metric, which we demonstrate using the EDGB and dilaton metrics. We compare the KRZ metric with the bumpy metric, focusing on the $\delta_1$ and $\delta_6$ parameters for simplicity. The bumpy black hole metric has a multipolar structure that closely resembles, but is not exactly, that of a black hole. Notably, a clear correlation between $\delta_1$, $\delta_6$, and the quadrupole moment $Q$ is identified. 

In Sec~\ref{Inspiral}, we analyze the inspiral waveform phase in the KRZ metric, exploring the impact of varying parameters. The ringdown waveform is derived from the properties of the photon sphere\cite{Yang_12} around black holes \cite{BOB_b,Psi_22}. This section gives a brief overview of how to obtain quasinormal modes using the photon sphere and how to derive the ringdown waveform from these modes. Next, we establish a connection between the inspiral and ringdown waveforms, considering the peak as the matching point. We refer to this full waveform model as $\Psi_{\mathrm{FD}}$, because this waveform model depends on the specific metric, it is difficult to apply it to an unknown metric. We plot the full waveform with different spins $a$ and different quadrupole moment deviations $\Delta Q$ and find the quadrupole moment deviation $\Delta Q$ has a significant influence on the inspiral waveform. 

We also calculate the overlap between the waveforms from KRZ and Kerr binary black holes. If the SNR is enough (roughly around 40), LIGO-Virgo-Kagra (LVK) may constrain the deviation of a quadrupole moment from the Kerr black hole in a relative error $\sim 10^{-3}$. However, in the previous detections by LVK, there is no event with such a high SNR. We may participate in the O4 run, and LVK will find candidates with higher SNRs. For the next-generation detectors such as ET and space-borne detectors Taiji and LISA, GW signals with larger SNRs could easily be detected; therefore, the next generation will have a better measurement of the deviation from Kerr black holes. 

We also found that the effect of deviation $\Delta Q$ on the waveforms has a relation with the spin parameter $a$. Our findings revealed that as the spin increases, the influence of $\Delta Q/Q_{\mathrm{Kerr}}$ diminishes. This trend can be explained by considering the definition of the quadrupole moment $Q$ in the KRZ metric. We believe that our waveform model can be useful for testing the no-hair theorem with GWs. In an upcoming work, our waveform template will be employed to conduct data analysis to test non-GR black holes with LVK events. This will provide insights into their properties and offer a potential avenue for future tests of general relativity.

\section*{ACKNOWLEDGMENTS}
This work is supported by The National Key R\&D Program
of China (Grant No. 2021YFC2203002), NSFC (National Natural Science Foundation of China) Grants No. 11773059 and No. 12173071. W. H. is supported by CAS Project for Young Scientists in Basic Research YSBR-006.

\bibliographystyle{apsrev4-1}  %% BibTeX style

\begin{thebibliography}{89}%
\makeatletter
\providecommand \@ifxundefined [1]{%
 \@ifx{#1\undefined}
}%
\providecommand \@ifnum [1]{%
 \ifnum #1\expandafter \@firstoftwo
 \else \expandafter \@secondoftwo
 \fi
}%
\providecommand \@ifx [1]{%
 \ifx #1\expandafter \@firstoftwo
 \else \expandafter \@secondoftwo
 \fi
}%
\providecommand \natexlab [1]{#1}%
\providecommand \enquote  [1]{``#1''}%
\providecommand \bibnamefont  [1]{#1}%
\providecommand \bibfnamefont [1]{#1}%
\providecommand \citenamefont [1]{#1}%
\providecommand \href@noop [0]{\@secondoftwo}%
\providecommand \href [0]{\begingroup \@sanitize@url \@href}%
\providecommand \@href[1]{\@@startlink{#1}\@@href}%
\providecommand \@@href[1]{\endgroup#1\@@endlink}%
\providecommand \@sanitize@url [0]{\catcode `\\12\catcode `\$12\catcode
  `\&12\catcode `\#12\catcode `\^12\catcode `\_12\catcode `\%12\relax}%
\providecommand \@@startlink[1]{}%
\providecommand \@@endlink[0]{}%
\providecommand \url  [0]{\begingroup\@sanitize@url \@url }%
\providecommand \@url [1]{\endgroup\@href {#1}{\urlprefix }}%
\providecommand \urlprefix  [0]{URL }%
\providecommand \Eprint [0]{\href }%
\providecommand \doibase [0]{http://dx.doi.org/}%
\providecommand \selectlanguage [0]{\@gobble}%
\providecommand \bibinfo  [0]{\@secondoftwo}%
\providecommand \bibfield  [0]{\@secondoftwo}%
\providecommand \translation [1]{[#1]}%
\providecommand \BibitemOpen [0]{}%
\providecommand \bibitemStop [0]{}%
\providecommand \bibitemNoStop [0]{.\EOS\space}%
\providecommand \EOS [0]{\spacefactor3000\relax}%
\providecommand \BibitemShut  [1]{\csname bibitem#1\endcsname}%
\let\auto@bib@innerbib\@empty
%</preamble>
\bibitem [{\citenamefont {Abbott}\ \emph
  {et~al.}(2016{\natexlab{a}})\citenamefont {Abbott} \emph
  {et~al.}}]{GW150914}%
  \BibitemOpen
  \bibfield  {author} {\bibinfo {author} {\bibfnamefont {B.~P.}\ \bibnamefont
  {Abbott}} \emph {et~al.} (\bibinfo {collaboration} {LIGO Scientific
  Collaboration and Virgo Collaboration}),\ }\href {\doibase
  10.1103/PhysRevLett.116.061102} {\bibfield  {journal} {\bibinfo  {journal}
  {Phys. Rev. Lett.}\ }\textbf {\bibinfo {volume} {116}},\ \bibinfo {pages}
  {061102} (\bibinfo {year} {2016}{\natexlab{a}})}\BibitemShut {NoStop}%
\bibitem [{\citenamefont {Abbott}\ \emph
  {et~al.}(2016{\natexlab{b}})\citenamefont {Abbott} \emph
  {et~al.}}]{GW_events_1}%
  \BibitemOpen
  \bibfield  {author} {\bibinfo {author} {\bibfnamefont {B.~P.}\ \bibnamefont
  {Abbott}} \emph {et~al.} (\bibinfo {collaboration} {LIGO Scientific
  Collaboration and Virgo Collaboration}),\ }\href {\doibase
  10.1103/PhysRevLett.116.241103} {\bibfield  {journal} {\bibinfo  {journal}
  {Phys. Rev. Lett.}\ }\textbf {\bibinfo {volume} {116}},\ \bibinfo {pages}
  {241103} (\bibinfo {year} {2016}{\natexlab{b}})}\BibitemShut {NoStop}%
\bibitem [{\citenamefont {Abbott}\ \emph
  {et~al.}(2019{\natexlab{a}})\citenamefont {Abbott} \emph
  {et~al.}}]{GW_events_2}%
  \BibitemOpen
  \bibfield  {author} {\bibinfo {author} {\bibfnamefont {B.~P.}\ \bibnamefont
  {Abbott}} \emph {et~al.} (\bibinfo {collaboration} {LIGO Scientific
  Collaboration and Virgo Collaboration}),\ }\href {\doibase
  10.1103/PhysRevX.9.031040} {\bibfield  {journal} {\bibinfo  {journal} {Phys.
  Rev. X}\ }\textbf {\bibinfo {volume} {9}},\ \bibinfo {pages} {031040}
  (\bibinfo {year} {2019}{\natexlab{a}})}\BibitemShut {NoStop}%
\bibitem [{\citenamefont {Venumadhav}\ \emph {et~al.}(2020)\citenamefont
  {Venumadhav}, \citenamefont {Zackay}, \citenamefont {Roulet}, \citenamefont
  {Dai},\ and\ \citenamefont {Zaldarriaga}}]{GW_events_3}%
  \BibitemOpen
  \bibfield  {author} {\bibinfo {author} {\bibfnamefont {T.}~\bibnamefont
  {Venumadhav}}, \bibinfo {author} {\bibfnamefont {B.}~\bibnamefont {Zackay}},
  \bibinfo {author} {\bibfnamefont {J.}~\bibnamefont {Roulet}}, \bibinfo
  {author} {\bibfnamefont {L.}~\bibnamefont {Dai}}, \ and\ \bibinfo {author}
  {\bibfnamefont {M.}~\bibnamefont {Zaldarriaga}},\ }\href {\doibase
  10.1103/PhysRevD.101.083030} {\bibfield  {journal} {\bibinfo  {journal}
  {Phys. Rev. D}\ }\textbf {\bibinfo {volume} {101}},\ \bibinfo {pages}
  {083030} (\bibinfo {year} {2020})}\BibitemShut {NoStop}%
\bibitem [{\citenamefont {Abbott}\ \emph
  {et~al.}(2021{\natexlab{a}})\citenamefont {Abbott} \emph
  {et~al.}}]{GW_events_4}%
  \BibitemOpen
  \bibfield  {author} {\bibinfo {author} {\bibfnamefont {R.}~\bibnamefont
  {Abbott}} \emph {et~al.} (\bibinfo {collaboration} {LIGO Scientific
  Collaboration and Virgo Collaboration}),\ }\href {\doibase
  10.1103/PhysRevX.11.021053} {\bibfield  {journal} {\bibinfo  {journal} {Phys.
  Rev. X}\ }\textbf {\bibinfo {volume} {11}},\ \bibinfo {pages} {021053}
  (\bibinfo {year} {2021}{\natexlab{a}})}\BibitemShut {NoStop}%
\bibitem [{\citenamefont {Abbott}\ \emph
  {et~al.}(2021{\natexlab{b}})\citenamefont {Abbott} \emph {et~al.}}]{BN}%
  \BibitemOpen
  \bibfield  {author} {\bibinfo {author} {\bibfnamefont {R.}~\bibnamefont
  {Abbott}} \emph {et~al.},\ }\href {\doibase 10.3847/2041-8213/ac082e}
  {\bibfield  {journal} {\bibinfo  {journal} {Astrophys. J. Lett}\ }\textbf
  {\bibinfo {volume} {915}},\ \bibinfo {pages} {L5} (\bibinfo {year}
  {2021}{\natexlab{b}})}\BibitemShut {NoStop}%
\bibitem [{\citenamefont {Okounkova}(2020)}]{Test_GR_1}%
  \BibitemOpen
  \bibfield  {author} {\bibinfo {author} {\bibfnamefont {M.}~\bibnamefont
  {Okounkova}},\ }\href {\doibase 10.1103/PhysRevD.102.084046} {\bibfield
  {journal} {\bibinfo  {journal} {Phys. Rev. D}\ }\textbf {\bibinfo {volume}
  {102}},\ \bibinfo {pages} {084046} (\bibinfo {year} {2020})}\BibitemShut
  {NoStop}%
\bibitem [{\citenamefont {Isi}\ \emph {et~al.}(2019)\citenamefont {Isi},
  \citenamefont {Giesler}, \citenamefont {Farr}, \citenamefont {Scheel},\ and\
  \citenamefont {Teukolsky}}]{Test_GR_2}%
  \BibitemOpen
  \bibfield  {author} {\bibinfo {author} {\bibfnamefont {M.}~\bibnamefont
  {Isi}}, \bibinfo {author} {\bibfnamefont {M.}~\bibnamefont {Giesler}},
  \bibinfo {author} {\bibfnamefont {W.~M.}\ \bibnamefont {Farr}}, \bibinfo
  {author} {\bibfnamefont {M.~A.}\ \bibnamefont {Scheel}}, \ and\ \bibinfo
  {author} {\bibfnamefont {S.~A.}\ \bibnamefont {Teukolsky}},\ }\href {\doibase
  10.1103/PhysRevLett.123.111102} {\bibfield  {journal} {\bibinfo  {journal}
  {Phys. Rev. Lett.}\ }\textbf {\bibinfo {volume} {123}},\ \bibinfo {pages}
  {111102} (\bibinfo {year} {2019})}\BibitemShut {NoStop}%
\bibitem [{\citenamefont {Abbott}\ \emph
  {et~al.}(2019{\natexlab{b}})\citenamefont {Abbott} \emph
  {et~al.}}]{Test_GR_3}%
  \BibitemOpen
  \bibfield  {author} {\bibinfo {author} {\bibfnamefont {B.~P.}\ \bibnamefont
  {Abbott}} \emph {et~al.} (\bibinfo {collaboration} {LIGO Scientific
  Collaboration and Virgo Collaboration}),\ }\href {\doibase
  10.1103/PhysRevLett.123.011102} {\bibfield  {journal} {\bibinfo  {journal}
  {Phys. Rev. Lett.}\ }\textbf {\bibinfo {volume} {123}},\ \bibinfo {pages}
  {011102} (\bibinfo {year} {2019}{\natexlab{b}})}\BibitemShut {NoStop}%
\bibitem [{\citenamefont {Nair}\ \emph {et~al.}(2019)\citenamefont {Nair},
  \citenamefont {Perkins}, \citenamefont {Silva},\ and\ \citenamefont
  {Yunes}}]{Test_GR_4}%
  \BibitemOpen
  \bibfield  {author} {\bibinfo {author} {\bibfnamefont {R.}~\bibnamefont
  {Nair}}, \bibinfo {author} {\bibfnamefont {S.}~\bibnamefont {Perkins}},
  \bibinfo {author} {\bibfnamefont {H.~O.}\ \bibnamefont {Silva}}, \ and\
  \bibinfo {author} {\bibfnamefont {N.}~\bibnamefont {Yunes}},\ }\href
  {\doibase 10.1103/PhysRevLett.123.191101} {\bibfield  {journal} {\bibinfo
  {journal} {Phys. Rev. Lett.}\ }\textbf {\bibinfo {volume} {123}},\ \bibinfo
  {pages} {191101} (\bibinfo {year} {2019})}\BibitemShut {NoStop}%
\bibitem [{\citenamefont {Abbott}\ \emph
  {et~al.}(2019{\natexlab{c}})\citenamefont {Abbott} \emph
  {et~al.}}]{Test_GR_5}%
  \BibitemOpen
  \bibfield  {author} {\bibinfo {author} {\bibfnamefont {B.~P.}\ \bibnamefont
  {Abbott}} \emph {et~al.} (\bibinfo {collaboration} {The LIGO Scientific
  Collaboration and the Virgo Collaboration}),\ }\href {\doibase
  10.1103/PhysRevD.100.104036} {\bibfield  {journal} {\bibinfo  {journal}
  {Phys. Rev. D}\ }\textbf {\bibinfo {volume} {100}},\ \bibinfo {pages}
  {104036} (\bibinfo {year} {2019}{\natexlab{c}})}\BibitemShut {NoStop}%
\bibitem [{\citenamefont {Abbott}\ \emph
  {et~al.}(2021{\natexlab{c}})\citenamefont {Abbott} \emph
  {et~al.}}]{Test_GR_6}%
  \BibitemOpen
  \bibfield  {author} {\bibinfo {author} {\bibfnamefont {R.}~\bibnamefont
  {Abbott}} \emph {et~al.} (\bibinfo {collaboration} {LIGO Scientific
  Collaboration and Virgo Collaboration}),\ }\href {\doibase
  10.1103/PhysRevD.103.122002} {\bibfield  {journal} {\bibinfo  {journal}
  {Phys. Rev. D}\ }\textbf {\bibinfo {volume} {103}},\ \bibinfo {pages}
  {122002} (\bibinfo {year} {2021}{\natexlab{c}})}\BibitemShut {NoStop}%
\bibitem [{\citenamefont {Abbott}\ \emph
  {et~al.}(2019{\natexlab{d}})\citenamefont {Abbott} \emph {et~al.}}]{CO_1}%
  \BibitemOpen
  \bibfield  {author} {\bibinfo {author} {\bibfnamefont {B.~P.}\ \bibnamefont
  {Abbott}} \emph {et~al.},\ }\href {\doibase 10.3847/2041-8213/ab3800}
  {\bibfield  {journal} {\bibinfo  {journal} {The Astrophysical Journal
  Letters}\ }\textbf {\bibinfo {volume} {882}},\ \bibinfo {pages} {L24}
  (\bibinfo {year} {2019}{\natexlab{d}})}\BibitemShut {NoStop}%
\bibitem [{\citenamefont {{Huang}}\ \emph {et~al.}(2017)\citenamefont
  {{Huang}}, \citenamefont {{Gong}}, \citenamefont {{Xu}}, \citenamefont
  {{Amaro-Seoane}}, \citenamefont {{Bian}}, \citenamefont {{Chen}},
  \citenamefont {{Chen}}, \citenamefont {{Fang}}, \citenamefont {{Feng}},
  \citenamefont {{Liu}}, \citenamefont {{Li}}, \citenamefont {{Li}},
  \citenamefont {{Luo}}, \citenamefont {{Shao}}, \citenamefont {{Spurzem}},
  \citenamefont {{Tang}}, \citenamefont {{Wang}}, \citenamefont {{Wang}},
  \citenamefont {{Zang}},\ and\ \citenamefont {{Lau}}}]{taiji}%
  \BibitemOpen
  \bibfield  {author} {\bibinfo {author} {\bibfnamefont {S.}~\bibnamefont
  {{Huang}}}, \bibinfo {author} {\bibfnamefont {X.}~\bibnamefont {{Gong}}},
  \bibinfo {author} {\bibfnamefont {P.}~\bibnamefont {{Xu}}}, \bibinfo {author}
  {\bibfnamefont {P.}~\bibnamefont {{Amaro-Seoane}}}, \bibinfo {author}
  {\bibfnamefont {X.}~\bibnamefont {{Bian}}}, \bibinfo {author} {\bibfnamefont
  {Y.}~\bibnamefont {{Chen}}}, \bibinfo {author} {\bibfnamefont
  {X.}~\bibnamefont {{Chen}}}, \bibinfo {author} {\bibfnamefont
  {Z.}~\bibnamefont {{Fang}}}, \bibinfo {author} {\bibfnamefont
  {X.}~\bibnamefont {{Feng}}}, \bibinfo {author} {\bibfnamefont
  {F.}~\bibnamefont {{Liu}}}, \bibinfo {author} {\bibfnamefont
  {S.}~\bibnamefont {{Li}}}, \bibinfo {author} {\bibfnamefont {X.}~\bibnamefont
  {{Li}}}, \bibinfo {author} {\bibfnamefont {Z.}~\bibnamefont {{Luo}}},
  \bibinfo {author} {\bibfnamefont {M.}~\bibnamefont {{Shao}}}, \bibinfo
  {author} {\bibfnamefont {R.}~\bibnamefont {{Spurzem}}}, \bibinfo {author}
  {\bibfnamefont {W.}~\bibnamefont {{Tang}}}, \bibinfo {author} {\bibfnamefont
  {Y.}~\bibnamefont {{Wang}}}, \bibinfo {author} {\bibfnamefont
  {Y.}~\bibnamefont {{Wang}}}, \bibinfo {author} {\bibfnamefont
  {Y.}~\bibnamefont {{Zang}}}, \ and\ \bibinfo {author} {\bibfnamefont
  {Y.}~\bibnamefont {{Lau}}},\ }\href {\doibase 10.1360/SSPMA2016-00438}
  {\bibfield  {journal} {\bibinfo  {journal} {Scientia Sinica Physica,
  Mechanica \& Astronomica}\ }\textbf {\bibinfo {volume} {47}},\ \bibinfo
  {pages} {010404} (\bibinfo {year} {2017})}\BibitemShut {NoStop}%
\bibitem [{\citenamefont {{Luo}}\ \emph {et~al.}(2016)\citenamefont {{Luo}},
  \citenamefont {{Chen}}, \citenamefont {{Duan}}, \citenamefont {{Gong}},
  \citenamefont {{Hu}}, \citenamefont {{Ji}}, \citenamefont {{Liu}},
  \citenamefont {{Mei}}, \citenamefont {{Milyukov}}, \citenamefont {{Sazhin}},
  \citenamefont {{Shao}}, \citenamefont {{Toth}}, \citenamefont {{Tu}},
  \citenamefont {{Wang}}, \citenamefont {{Wang}}, \citenamefont {{Yeh}},
  \citenamefont {{Zhan}}, \citenamefont {{Zhang}}, \citenamefont {{Zharov}},\
  and\ \citenamefont {{Zhou}}}]{tianqin}%
  \BibitemOpen
  \bibfield  {author} {\bibinfo {author} {\bibfnamefont {J.}~\bibnamefont
  {{Luo}}}, \bibinfo {author} {\bibfnamefont {L.-S.}\ \bibnamefont {{Chen}}},
  \bibinfo {author} {\bibfnamefont {H.-Z.}\ \bibnamefont {{Duan}}}, \bibinfo
  {author} {\bibfnamefont {Y.-G.}\ \bibnamefont {{Gong}}}, \bibinfo {author}
  {\bibfnamefont {S.}~\bibnamefont {{Hu}}}, \bibinfo {author} {\bibfnamefont
  {J.}~\bibnamefont {{Ji}}}, \bibinfo {author} {\bibfnamefont {Q.}~\bibnamefont
  {{Liu}}}, \bibinfo {author} {\bibfnamefont {J.}~\bibnamefont {{Mei}}},
  \bibinfo {author} {\bibfnamefont {V.}~\bibnamefont {{Milyukov}}}, \bibinfo
  {author} {\bibfnamefont {M.}~\bibnamefont {{Sazhin}}}, \bibinfo {author}
  {\bibfnamefont {C.-G.}\ \bibnamefont {{Shao}}}, \bibinfo {author}
  {\bibfnamefont {V.~T.}\ \bibnamefont {{Toth}}}, \bibinfo {author}
  {\bibfnamefont {H.-B.}\ \bibnamefont {{Tu}}}, \bibinfo {author}
  {\bibfnamefont {Y.}~\bibnamefont {{Wang}}}, \bibinfo {author} {\bibfnamefont
  {Y.}~\bibnamefont {{Wang}}}, \bibinfo {author} {\bibfnamefont {H.-C.}\
  \bibnamefont {{Yeh}}}, \bibinfo {author} {\bibfnamefont {M.-S.}\ \bibnamefont
  {{Zhan}}}, \bibinfo {author} {\bibfnamefont {Y.}~\bibnamefont {{Zhang}}},
  \bibinfo {author} {\bibfnamefont {V.}~\bibnamefont {{Zharov}}}, \ and\
  \bibinfo {author} {\bibfnamefont {Z.-B.}\ \bibnamefont {{Zhou}}},\ }\href
  {\doibase 10.1088/0264-9381/33/3/035010} {\bibfield  {journal} {\bibinfo
  {journal} {Classical and Quantum Gravity}\ }\textbf {\bibinfo {volume}
  {33}},\ \bibinfo {eid} {035010} (\bibinfo {year} {2016})}\BibitemShut
  {NoStop}%
\bibitem [{\citenamefont {Gair}\ \emph {et~al.}(2008)\citenamefont {Gair},
  \citenamefont {Li},\ and\ \citenamefont {Mandel}}]{Ricci_1}%
  \BibitemOpen
  \bibfield  {author} {\bibinfo {author} {\bibfnamefont {J.~R.}\ \bibnamefont
  {Gair}}, \bibinfo {author} {\bibfnamefont {C.}~\bibnamefont {Li}}, \ and\
  \bibinfo {author} {\bibfnamefont {I.}~\bibnamefont {Mandel}},\ }\href
  {\doibase 10.1103/PhysRevD.77.024035} {\bibfield  {journal} {\bibinfo
  {journal} {Phys. Rev. D}\ }\textbf {\bibinfo {volume} {77}},\ \bibinfo
  {pages} {024035} (\bibinfo {year} {2008})}\BibitemShut {NoStop}%
\bibitem [{\citenamefont {Johannsen}(2013)}]{Ricci_2}%
  \BibitemOpen
  \bibfield  {author} {\bibinfo {author} {\bibfnamefont {T.}~\bibnamefont
  {Johannsen}},\ }\href {\doibase 10.1103/PhysRevD.87.124017} {\bibfield
  {journal} {\bibinfo  {journal} {Phys. Rev. D}\ }\textbf {\bibinfo {volume}
  {87}},\ \bibinfo {pages} {124017} (\bibinfo {year} {2013})}\BibitemShut
  {NoStop}%
\bibitem [{\citenamefont {{Manko}}\ and\ \citenamefont
  {{Novikov}}(1992)}]{Ricci_3}%
  \BibitemOpen
  \bibfield  {author} {\bibinfo {author} {\bibfnamefont {V.~S.}\ \bibnamefont
  {{Manko}}}\ and\ \bibinfo {author} {\bibfnamefont {I.~D.}\ \bibnamefont
  {{Novikov}}},\ }\href {\doibase 10.1088/0264-9381/9/11/013} {\bibfield
  {journal} {\bibinfo  {journal} {Classical and Quantum Gravity}\ }\textbf
  {\bibinfo {volume} {9}},\ \bibinfo {pages} {2477} (\bibinfo {year}
  {1992})}\BibitemShut {NoStop}%
\bibitem [{\citenamefont {Papadopoulos}\ \emph {et~al.}(1981)\citenamefont
  {Papadopoulos}, \citenamefont {Stewart},\ and\ \citenamefont
  {Witten}}]{ZV_1}%
  \BibitemOpen
  \bibfield  {author} {\bibinfo {author} {\bibfnamefont {D.}~\bibnamefont
  {Papadopoulos}}, \bibinfo {author} {\bibfnamefont {B.}~\bibnamefont
  {Stewart}}, \ and\ \bibinfo {author} {\bibfnamefont {L.}~\bibnamefont
  {Witten}},\ }\href {\doibase 10.1103/PhysRevD.24.320} {\bibfield  {journal}
  {\bibinfo  {journal} {Phys. Rev. D}\ }\textbf {\bibinfo {volume} {24}},\
  \bibinfo {pages} {320} (\bibinfo {year} {1981})}\BibitemShut {NoStop}%
\bibitem [{\citenamefont {Chowdhury}\ \emph {et~al.}(2012)\citenamefont
  {Chowdhury}, \citenamefont {Patil}, \citenamefont {Malafarina},\ and\
  \citenamefont {Joshi}}]{ZV_2}%
  \BibitemOpen
  \bibfield  {author} {\bibinfo {author} {\bibfnamefont {A.~N.}\ \bibnamefont
  {Chowdhury}}, \bibinfo {author} {\bibfnamefont {M.}~\bibnamefont {Patil}},
  \bibinfo {author} {\bibfnamefont {D.}~\bibnamefont {Malafarina}}, \ and\
  \bibinfo {author} {\bibfnamefont {P.~S.}\ \bibnamefont {Joshi}},\ }\href
  {\doibase 10.1103/PhysRevD.85.104031} {\bibfield  {journal} {\bibinfo
  {journal} {Phys. Rev. D}\ }\textbf {\bibinfo {volume} {85}},\ \bibinfo
  {pages} {104031} (\bibinfo {year} {2012})}\BibitemShut {NoStop}%
\bibitem [{\citenamefont {Boshkayev}\ \emph {et~al.}(2016)\citenamefont
  {Boshkayev}, \citenamefont {Gasper\'{\i}n}, \citenamefont
  {Guti\'errez-Pi\~neres}, \citenamefont {Quevedo},\ and\ \citenamefont
  {Toktarbay}}]{ZV_3}%
  \BibitemOpen
  \bibfield  {author} {\bibinfo {author} {\bibfnamefont {K.}~\bibnamefont
  {Boshkayev}}, \bibinfo {author} {\bibfnamefont {E.}~\bibnamefont
  {Gasper\'{\i}n}}, \bibinfo {author} {\bibfnamefont {A.~C.}\ \bibnamefont
  {Guti\'errez-Pi\~neres}}, \bibinfo {author} {\bibfnamefont {H.}~\bibnamefont
  {Quevedo}}, \ and\ \bibinfo {author} {\bibfnamefont {S.}~\bibnamefont
  {Toktarbay}},\ }\href {\doibase 10.1103/PhysRevD.93.024024} {\bibfield
  {journal} {\bibinfo  {journal} {Phys. Rev. D}\ }\textbf {\bibinfo {volume}
  {93}},\ \bibinfo {pages} {024024} (\bibinfo {year} {2016})}\BibitemShut
  {NoStop}%
\bibitem [{\citenamefont {Toshmatov}\ \emph {et~al.}(2019)\citenamefont
  {Toshmatov}, \citenamefont {Malafarina},\ and\ \citenamefont
  {Dadhich}}]{ZV_4}%
  \BibitemOpen
  \bibfield  {author} {\bibinfo {author} {\bibfnamefont {B.}~\bibnamefont
  {Toshmatov}}, \bibinfo {author} {\bibfnamefont {D.}~\bibnamefont
  {Malafarina}}, \ and\ \bibinfo {author} {\bibfnamefont {N.}~\bibnamefont
  {Dadhich}},\ }\href {\doibase 10.1103/PhysRevD.100.044001} {\bibfield
  {journal} {\bibinfo  {journal} {Phys. Rev. D}\ }\textbf {\bibinfo {volume}
  {100}},\ \bibinfo {pages} {044001} (\bibinfo {year} {2019})}\BibitemShut
  {NoStop}%
\bibitem [{\citenamefont {Toshmatov}\ and\ \citenamefont
  {Malafarina}(2019)}]{ZV_5}%
  \BibitemOpen
  \bibfield  {author} {\bibinfo {author} {\bibfnamefont {B.}~\bibnamefont
  {Toshmatov}}\ and\ \bibinfo {author} {\bibfnamefont {D.}~\bibnamefont
  {Malafarina}},\ }\href {\doibase 10.1103/PhysRevD.100.104052} {\bibfield
  {journal} {\bibinfo  {journal} {Phys. Rev. D}\ }\textbf {\bibinfo {volume}
  {100}},\ \bibinfo {pages} {104052} (\bibinfo {year} {2019})}\BibitemShut
  {NoStop}%
\bibitem [{\citenamefont {Quevedo}\ and\ \citenamefont
  {Mashhoon}(1991)}]{deltaKerr_1}%
  \BibitemOpen
  \bibfield  {author} {\bibinfo {author} {\bibfnamefont {H.}~\bibnamefont
  {Quevedo}}\ and\ \bibinfo {author} {\bibfnamefont {B.}~\bibnamefont
  {Mashhoon}},\ }\href {\doibase 10.1103/PhysRevD.43.3902} {\bibfield
  {journal} {\bibinfo  {journal} {Phys. Rev. D}\ }\textbf {\bibinfo {volume}
  {43}},\ \bibinfo {pages} {3902} (\bibinfo {year} {1991})}\BibitemShut
  {NoStop}%
\bibitem [{\citenamefont {{Allahyari}}\ \emph {et~al.}(2020)\citenamefont
  {{Allahyari}}, \citenamefont {{Firouzjahi}},\ and\ \citenamefont
  {{Mashhoon}}}]{deltaKerr_2}%
  \BibitemOpen
  \bibfield  {author} {\bibinfo {author} {\bibfnamefont {A.}~\bibnamefont
  {{Allahyari}}}, \bibinfo {author} {\bibfnamefont {H.}~\bibnamefont
  {{Firouzjahi}}}, \ and\ \bibinfo {author} {\bibfnamefont {B.}~\bibnamefont
  {{Mashhoon}}},\ }\href {\doibase 10.1088/1361-6382/ab6860} {\bibfield
  {journal} {\bibinfo  {journal} {Classical and Quantum Gravity}\ }\textbf
  {\bibinfo {volume} {37}},\ \bibinfo {eid} {055006} (\bibinfo {year}
  {2020})}\BibitemShut {NoStop}%
\bibitem [{\citenamefont {Toktarbay}\ and\ \citenamefont
  {Quevedo}(2014)}]{deltaKerr_3}%
  \BibitemOpen
  \bibfield  {author} {\bibinfo {author} {\bibfnamefont {S.}~\bibnamefont
  {Toktarbay}}\ and\ \bibinfo {author} {\bibfnamefont {H.}~\bibnamefont
  {Quevedo}},\ }\href {\doibase 10.1134/S0202289314040136} {\bibfield
  {journal} {\bibinfo  {journal} {Gravitation and Cosmology}\ }\textbf
  {\bibinfo {volume} {20}},\ \bibinfo {pages} {252} (\bibinfo {year}
  {2014})}\BibitemShut {NoStop}%
\bibitem [{\citenamefont {Konoplya}\ \emph {et~al.}(2016)\citenamefont
  {Konoplya}, \citenamefont {Rezzolla},\ and\ \citenamefont
  {Zhidenko}}]{KRZ_16}%
  \BibitemOpen
  \bibfield  {author} {\bibinfo {author} {\bibfnamefont {R.}~\bibnamefont
  {Konoplya}}, \bibinfo {author} {\bibfnamefont {L.}~\bibnamefont {Rezzolla}},
  \ and\ \bibinfo {author} {\bibfnamefont {A.}~\bibnamefont {Zhidenko}},\
  }\href {\doibase 10.1103/PhysRevD.93.064015} {\bibfield  {journal} {\bibinfo
  {journal} {Phys. Rev. D}\ }\textbf {\bibinfo {volume} {93}},\ \bibinfo
  {pages} {064015} (\bibinfo {year} {2016})}\BibitemShut {NoStop}%
\bibitem [{\citenamefont {Akiyama}\ \emph
  {et~al.}(2019{\natexlab{a}})\citenamefont {Akiyama} \emph {et~al.}}]{EHT_1}%
  \BibitemOpen
  \bibfield  {author} {\bibinfo {author} {\bibfnamefont {K.}~\bibnamefont
  {Akiyama}} \emph {et~al.} (\bibinfo {collaboration} {The Event Horizon
  Telescope Collaboration}),\ }\href {\doibase 10.3847/2041-8213/ab0ec7}
  {\bibfield  {journal} {\bibinfo  {journal} {The Astrophysical Journal
  Letters}\ }\textbf {\bibinfo {volume} {875}},\ \bibinfo {pages} {L1}
  (\bibinfo {year} {2019}{\natexlab{a}})}\BibitemShut {NoStop}%
\bibitem [{\citenamefont {Akiyama}\ \emph
  {et~al.}(2019{\natexlab{b}})\citenamefont {Akiyama} \emph {et~al.}}]{EHT_2}%
  \BibitemOpen
  \bibfield  {author} {\bibinfo {author} {\bibfnamefont {K.}~\bibnamefont
  {Akiyama}} \emph {et~al.} (\bibinfo {collaboration} {The Event Horizon
  Telescope Collaboration}),\ }\href {\doibase 10.3847/2041-8213/ab0c96}
  {\bibfield  {journal} {\bibinfo  {journal} {The Astrophysical Journal
  Letters}\ }\textbf {\bibinfo {volume} {875}},\ \bibinfo {pages} {L2}
  (\bibinfo {year} {2019}{\natexlab{b}})}\BibitemShut {NoStop}%
\bibitem [{\citenamefont {Akiyama}\ \emph
  {et~al.}(2019{\natexlab{c}})\citenamefont {Akiyama} \emph {et~al.}}]{EHT_3}%
  \BibitemOpen
  \bibfield  {author} {\bibinfo {author} {\bibfnamefont {K.}~\bibnamefont
  {Akiyama}} \emph {et~al.} (\bibinfo {collaboration} {The Event Horizon
  Telescope Collaboration}),\ }\href {\doibase 10.3847/2041-8213/ab0c57}
  {\bibfield  {journal} {\bibinfo  {journal} {The Astrophysical Journal
  Letters}\ }\textbf {\bibinfo {volume} {875}},\ \bibinfo {pages} {L3}
  (\bibinfo {year} {2019}{\natexlab{c}})}\BibitemShut {NoStop}%
\bibitem [{\citenamefont {Akiyama}\ \emph {et~al.}(2022)\citenamefont {Akiyama}
  \emph {et~al.}}]{EHT_S}%
  \BibitemOpen
  \bibfield  {author} {\bibinfo {author} {\bibfnamefont {K.}~\bibnamefont
  {Akiyama}} \emph {et~al.} (\bibinfo {collaboration} {The Event Horizon
  Telescope Collaboration}),\ }\href {\doibase 10.3847/2041-8213/ac6674}
  {\bibfield  {journal} {\bibinfo  {journal} {The Astrophysical Journal
  Letters}\ }\textbf {\bibinfo {volume} {930}},\ \bibinfo {pages} {L12}
  (\bibinfo {year} {2022})}\BibitemShut {NoStop}%
\bibitem [{\citenamefont {Gralla}(2021)}]{Test_GR_Sha_1}%
  \BibitemOpen
  \bibfield  {author} {\bibinfo {author} {\bibfnamefont {S.~E.}\ \bibnamefont
  {Gralla}},\ }\href {\doibase 10.1103/PhysRevD.103.024023} {\bibfield
  {journal} {\bibinfo  {journal} {Phys. Rev. D}\ }\textbf {\bibinfo {volume}
  {103}},\ \bibinfo {pages} {024023} (\bibinfo {year} {2021})}\BibitemShut
  {NoStop}%
\bibitem [{\citenamefont {Glampedakis}\ and\ \citenamefont
  {Pappas}(2021)}]{Test_GR_Sha_2}%
  \BibitemOpen
  \bibfield  {author} {\bibinfo {author} {\bibfnamefont {K.}~\bibnamefont
  {Glampedakis}}\ and\ \bibinfo {author} {\bibfnamefont {G.}~\bibnamefont
  {Pappas}},\ }\href {\doibase 10.1103/PhysRevD.104.L081503} {\bibfield
  {journal} {\bibinfo  {journal} {Phys. Rev. D}\ }\textbf {\bibinfo {volume}
  {104}},\ \bibinfo {pages} {L081503} (\bibinfo {year} {2021})}\BibitemShut
  {NoStop}%
\bibitem [{\citenamefont {Bambi}\ \emph {et~al.}(2019)\citenamefont {Bambi},
  \citenamefont {Freese}, \citenamefont {Vagnozzi},\ and\ \citenamefont
  {Visinelli}}]{Test_GR_Sha_3}%
  \BibitemOpen
  \bibfield  {author} {\bibinfo {author} {\bibfnamefont {C.}~\bibnamefont
  {Bambi}}, \bibinfo {author} {\bibfnamefont {K.}~\bibnamefont {Freese}},
  \bibinfo {author} {\bibfnamefont {S.}~\bibnamefont {Vagnozzi}}, \ and\
  \bibinfo {author} {\bibfnamefont {L.}~\bibnamefont {Visinelli}},\ }\href
  {\doibase 10.1103/PhysRevD.100.044057} {\bibfield  {journal} {\bibinfo
  {journal} {Phys. Rev. D}\ }\textbf {\bibinfo {volume} {100}},\ \bibinfo
  {pages} {044057} (\bibinfo {year} {2019})}\BibitemShut {NoStop}%
\bibitem [{\citenamefont {Virbhadra}(2009)}]{Light_Bend_1}%
  \BibitemOpen
  \bibfield  {author} {\bibinfo {author} {\bibfnamefont {K.~S.}\ \bibnamefont
  {Virbhadra}},\ }\href {\doibase 10.1103/PhysRevD.79.083004} {\bibfield
  {journal} {\bibinfo  {journal} {Phys. Rev. D}\ }\textbf {\bibinfo {volume}
  {79}},\ \bibinfo {pages} {083004} (\bibinfo {year} {2009})}\BibitemShut
  {NoStop}%
\bibitem [{\citenamefont {Virbhadra}\ and\ \citenamefont
  {Ellis}(2000)}]{Light_Bend_2}%
  \BibitemOpen
  \bibfield  {author} {\bibinfo {author} {\bibfnamefont {K.~S.}\ \bibnamefont
  {Virbhadra}}\ and\ \bibinfo {author} {\bibfnamefont {G.~F.~R.}\ \bibnamefont
  {Ellis}},\ }\href {\doibase 10.1103/PhysRevD.62.084003} {\bibfield  {journal}
  {\bibinfo  {journal} {Phys. Rev. D}\ }\textbf {\bibinfo {volume} {62}},\
  \bibinfo {pages} {084003} (\bibinfo {year} {2000})}\BibitemShut {NoStop}%
\bibitem [{\citenamefont {Bozza}(2002)}]{Light_Bend_3}%
  \BibitemOpen
  \bibfield  {author} {\bibinfo {author} {\bibfnamefont {V.}~\bibnamefont
  {Bozza}},\ }\href {\doibase 10.1103/PhysRevD.66.103001} {\bibfield  {journal}
  {\bibinfo  {journal} {Phys. Rev. D}\ }\textbf {\bibinfo {volume} {66}},\
  \bibinfo {pages} {103001} (\bibinfo {year} {2002})}\BibitemShut {NoStop}%
\bibitem [{\citenamefont {Gralla}\ \emph {et~al.}(2019)\citenamefont {Gralla},
  \citenamefont {Holz},\ and\ \citenamefont {Wald}}]{Light_Bend_4}%
  \BibitemOpen
  \bibfield  {author} {\bibinfo {author} {\bibfnamefont {S.~E.}\ \bibnamefont
  {Gralla}}, \bibinfo {author} {\bibfnamefont {D.~E.}\ \bibnamefont {Holz}}, \
  and\ \bibinfo {author} {\bibfnamefont {R.~M.}\ \bibnamefont {Wald}},\ }\href
  {\doibase 10.1103/PhysRevD.100.024018} {\bibfield  {journal} {\bibinfo
  {journal} {Phys. Rev. D}\ }\textbf {\bibinfo {volume} {100}},\ \bibinfo
  {pages} {024018} (\bibinfo {year} {2019})}\BibitemShut {NoStop}%
\bibitem [{\citenamefont {Bambi}\ and\ \citenamefont {Freese}(2009)}]{Image_1}%
  \BibitemOpen
  \bibfield  {author} {\bibinfo {author} {\bibfnamefont {C.}~\bibnamefont
  {Bambi}}\ and\ \bibinfo {author} {\bibfnamefont {K.}~\bibnamefont {Freese}},\
  }\href {\doibase 10.1103/PhysRevD.79.043002} {\bibfield  {journal} {\bibinfo
  {journal} {Phys. Rev. D}\ }\textbf {\bibinfo {volume} {79}},\ \bibinfo
  {pages} {043002} (\bibinfo {year} {2009})}\BibitemShut {NoStop}%
\bibitem [{\citenamefont {Hioki}\ and\ \citenamefont {Maeda}(2009)}]{Image_2}%
  \BibitemOpen
  \bibfield  {author} {\bibinfo {author} {\bibfnamefont {K.}~\bibnamefont
  {Hioki}}\ and\ \bibinfo {author} {\bibfnamefont {K.-i.}\ \bibnamefont
  {Maeda}},\ }\href {\doibase 10.1103/PhysRevD.80.024042} {\bibfield  {journal}
  {\bibinfo  {journal} {Phys. Rev. D}\ }\textbf {\bibinfo {volume} {80}},\
  \bibinfo {pages} {024042} (\bibinfo {year} {2009})}\BibitemShut {NoStop}%
\bibitem [{\citenamefont {Amarilla}\ \emph {et~al.}(2010)\citenamefont
  {Amarilla}, \citenamefont {Eiroa},\ and\ \citenamefont {Giribet}}]{Image_3}%
  \BibitemOpen
  \bibfield  {author} {\bibinfo {author} {\bibfnamefont {L.}~\bibnamefont
  {Amarilla}}, \bibinfo {author} {\bibfnamefont {E.~F.}\ \bibnamefont {Eiroa}},
  \ and\ \bibinfo {author} {\bibfnamefont {G.}~\bibnamefont {Giribet}},\ }\href
  {\doibase 10.1103/PhysRevD.81.124045} {\bibfield  {journal} {\bibinfo
  {journal} {Phys. Rev. D}\ }\textbf {\bibinfo {volume} {81}},\ \bibinfo
  {pages} {124045} (\bibinfo {year} {2010})}\BibitemShut {NoStop}%
\bibitem [{\citenamefont {Amarilla}\ and\ \citenamefont
  {Eiroa}(2012)}]{Image_4}%
  \BibitemOpen
  \bibfield  {author} {\bibinfo {author} {\bibfnamefont {L.}~\bibnamefont
  {Amarilla}}\ and\ \bibinfo {author} {\bibfnamefont {E.~F.}\ \bibnamefont
  {Eiroa}},\ }\href {\doibase 10.1103/PhysRevD.85.064019} {\bibfield  {journal}
  {\bibinfo  {journal} {Phys. Rev. D}\ }\textbf {\bibinfo {volume} {85}},\
  \bibinfo {pages} {064019} (\bibinfo {year} {2012})}\BibitemShut {NoStop}%
\bibitem [{\citenamefont {Amarilla}\ and\ \citenamefont
  {Eiroa}(2013)}]{Image_5}%
  \BibitemOpen
  \bibfield  {author} {\bibinfo {author} {\bibfnamefont {L.}~\bibnamefont
  {Amarilla}}\ and\ \bibinfo {author} {\bibfnamefont {E.~F.}\ \bibnamefont
  {Eiroa}},\ }\href {\doibase 10.1103/PhysRevD.87.044057} {\bibfield  {journal}
  {\bibinfo  {journal} {Phys. Rev. D}\ }\textbf {\bibinfo {volume} {87}},\
  \bibinfo {pages} {044057} (\bibinfo {year} {2013})}\BibitemShut {NoStop}%
\bibitem [{\citenamefont {Javed}\ \emph {et~al.}(2019)\citenamefont {Javed},
  \citenamefont {Abbas},\ and\ \citenamefont {\"Ovg\"un}}]{Image_6}%
  \BibitemOpen
  \bibfield  {author} {\bibinfo {author} {\bibfnamefont {W.}~\bibnamefont
  {Javed}}, \bibinfo {author} {\bibfnamefont {J.}~\bibnamefont {Abbas}}, \ and\
  \bibinfo {author} {\bibfnamefont {A.}~\bibnamefont {\"Ovg\"un}},\ }\href
  {\doibase 10.1103/PhysRevD.100.044052} {\bibfield  {journal} {\bibinfo
  {journal} {Phys. Rev. D}\ }\textbf {\bibinfo {volume} {100}},\ \bibinfo
  {pages} {044052} (\bibinfo {year} {2019})}\BibitemShut {NoStop}%
\bibitem [{\citenamefont {Younsi}\ \emph {et~al.}(2016)\citenamefont {Younsi},
  \citenamefont {Zhidenko}, \citenamefont {Rezzolla}, \citenamefont
  {Konoplya},\ and\ \citenamefont {Mizuno}}]{Image_7}%
  \BibitemOpen
  \bibfield  {author} {\bibinfo {author} {\bibfnamefont {Z.}~\bibnamefont
  {Younsi}}, \bibinfo {author} {\bibfnamefont {A.}~\bibnamefont {Zhidenko}},
  \bibinfo {author} {\bibfnamefont {L.}~\bibnamefont {Rezzolla}}, \bibinfo
  {author} {\bibfnamefont {R.}~\bibnamefont {Konoplya}}, \ and\ \bibinfo
  {author} {\bibfnamefont {Y.}~\bibnamefont {Mizuno}},\ }\href {\doibase
  10.1103/PhysRevD.94.084025} {\bibfield  {journal} {\bibinfo  {journal} {Phys.
  Rev. D}\ }\textbf {\bibinfo {volume} {94}},\ \bibinfo {pages} {084025}
  (\bibinfo {year} {2016})}\BibitemShut {NoStop}%
\bibitem [{\citenamefont {Cunha}\ \emph {et~al.}(2015)\citenamefont {Cunha},
  \citenamefont {Herdeiro}, \citenamefont {Radu},\ and\ \citenamefont
  {R\'unarsson}}]{Image_8}%
  \BibitemOpen
  \bibfield  {author} {\bibinfo {author} {\bibfnamefont {P.~V.~P.}\
  \bibnamefont {Cunha}}, \bibinfo {author} {\bibfnamefont {C.~A.~R.}\
  \bibnamefont {Herdeiro}}, \bibinfo {author} {\bibfnamefont {E.}~\bibnamefont
  {Radu}}, \ and\ \bibinfo {author} {\bibfnamefont {H.~F.}\ \bibnamefont
  {R\'unarsson}},\ }\href {\doibase 10.1103/PhysRevLett.115.211102} {\bibfield
  {journal} {\bibinfo  {journal} {Phys. Rev. Lett.}\ }\textbf {\bibinfo
  {volume} {115}},\ \bibinfo {pages} {211102} (\bibinfo {year}
  {2015})}\BibitemShut {NoStop}%
\bibitem [{\citenamefont {Ghasemi-Nodehi}\ \emph {et~al.}(2015)\citenamefont
  {Ghasemi-Nodehi}, \citenamefont {Li},\ and\ \citenamefont {Bambi}}]{Image_9}%
  \BibitemOpen
  \bibfield  {author} {\bibinfo {author} {\bibfnamefont {M.}~\bibnamefont
  {Ghasemi-Nodehi}}, \bibinfo {author} {\bibfnamefont {Z.}~\bibnamefont {Li}},
  \ and\ \bibinfo {author} {\bibfnamefont {C.}~\bibnamefont {Bambi}},\ }\href
  {\doibase 10.1140/epjc/s10052-015-3539-x} {\bibfield  {journal} {\bibinfo
  {journal} {The European Physical Journal C}\ }\textbf {\bibinfo {volume}
  {75}},\ \bibinfo {pages} {315} (\bibinfo {year} {2015})}\BibitemShut
  {NoStop}%
\bibitem [{\citenamefont {Ayzenberg}\ and\ \citenamefont
  {Yunes}(2018)}]{Image_10}%
  \BibitemOpen
  \bibfield  {author} {\bibinfo {author} {\bibfnamefont {D.}~\bibnamefont
  {Ayzenberg}}\ and\ \bibinfo {author} {\bibfnamefont {N.}~\bibnamefont
  {Yunes}},\ }\href {\doibase 10.1088/1361-6382/aae87b} {\bibfield  {journal}
  {\bibinfo  {journal} {Classical and Quantum Gravity}\ }\textbf {\bibinfo
  {volume} {35}},\ \bibinfo {pages} {235002} (\bibinfo {year}
  {2018})}\BibitemShut {NoStop}%
\bibitem [{\citenamefont {Claudel}\ \emph {et~al.}(2001)\citenamefont
  {Claudel}, \citenamefont {Virbhadra},\ and\ \citenamefont
  {Ellis}}]{Claude_01}%
  \BibitemOpen
  \bibfield  {author} {\bibinfo {author} {\bibfnamefont {C.-M.}\ \bibnamefont
  {Claudel}}, \bibinfo {author} {\bibfnamefont {K.~S.}\ \bibnamefont
  {Virbhadra}}, \ and\ \bibinfo {author} {\bibfnamefont {G.~F.~R.}\
  \bibnamefont {Ellis}},\ }\href {\doibase 10.1063/1.1308507} {\bibfield
  {journal} {\bibinfo  {journal} {Journal of Mathematical Physics}\ }\textbf
  {\bibinfo {volume} {42}},\ \bibinfo {pages} {818} (\bibinfo {year}
  {2001})}\BibitemShut {NoStop}%
\bibitem [{\citenamefont {Lu}\ \emph {et~al.}(2023)\citenamefont {Lu} \emph
  {et~al.}}]{Lu_23}%
  \BibitemOpen
  \bibfield  {author} {\bibinfo {author} {\bibfnamefont {R.-S.}\ \bibnamefont
  {Lu}} \emph {et~al.},\ }\href {\doibase 10.1038/s41586-023-05843-w}
  {\bibfield  {journal} {\bibinfo  {journal} {Nature}\ }\textbf {\bibinfo
  {volume} {616}},\ \bibinfo {pages} {686} (\bibinfo {year}
  {2023})}\BibitemShut {NoStop}%
\bibitem [{\citenamefont {Psaltis}\ and\ \citenamefont
  {Johannsen}(2011)}]{Ray_1}%
  \BibitemOpen
  \bibfield  {author} {\bibinfo {author} {\bibfnamefont {D.}~\bibnamefont
  {Psaltis}}\ and\ \bibinfo {author} {\bibfnamefont {T.}~\bibnamefont
  {Johannsen}},\ }\href {\doibase 10.1088/0004-637X/745/1/1} {\bibfield
  {journal} {\bibinfo  {journal} {The Astrophysical Journal}\ }\textbf
  {\bibinfo {volume} {745}},\ \bibinfo {pages} {1} (\bibinfo {year}
  {2011})}\BibitemShut {NoStop}%
\bibitem [{\citenamefont {kwan Chan}\ \emph {et~al.}(2013)\citenamefont {kwan
  Chan}, \citenamefont {Psaltis},\ and\ \citenamefont {Özel}}]{Ray_2}%
  \BibitemOpen
  \bibfield  {author} {\bibinfo {author} {\bibfnamefont {C.}~\bibnamefont {kwan
  Chan}}, \bibinfo {author} {\bibfnamefont {D.}~\bibnamefont {Psaltis}}, \ and\
  \bibinfo {author} {\bibfnamefont {F.}~\bibnamefont {Özel}},\ }\href
  {\doibase 10.1088/0004-637X/777/1/13} {\bibfield  {journal} {\bibinfo
  {journal} {The Astrophysical Journal}\ }\textbf {\bibinfo {volume} {777}},\
  \bibinfo {pages} {13} (\bibinfo {year} {2013})}\BibitemShut {NoStop}%
\bibitem [{\citenamefont {Pihajoki}\ \emph {et~al.}(2018)\citenamefont
  {Pihajoki}, \citenamefont {Mannerkoski}, \citenamefont {Nättilä},\ and\
  \citenamefont {Johansson}}]{Ray_3}%
  \BibitemOpen
  \bibfield  {author} {\bibinfo {author} {\bibfnamefont {P.}~\bibnamefont
  {Pihajoki}}, \bibinfo {author} {\bibfnamefont {M.}~\bibnamefont
  {Mannerkoski}}, \bibinfo {author} {\bibfnamefont {J.}~\bibnamefont
  {Nättilä}}, \ and\ \bibinfo {author} {\bibfnamefont {P.~H.}\ \bibnamefont
  {Johansson}},\ }\href {\doibase 10.3847/1538-4357/aacea0} {\bibfield
  {journal} {\bibinfo  {journal} {The Astrophysical Journal}\ }\textbf
  {\bibinfo {volume} {863}},\ \bibinfo {pages} {8} (\bibinfo {year}
  {2018})}\BibitemShut {NoStop}%
\bibitem [{\citenamefont {Pelle}\ \emph {et~al.}(2022)\citenamefont {Pelle},
  \citenamefont {Reula}, \citenamefont {Carrasco},\ and\ \citenamefont
  {Bederian}}]{Ray_4}%
  \BibitemOpen
  \bibfield  {author} {\bibinfo {author} {\bibfnamefont {J.}~\bibnamefont
  {Pelle}}, \bibinfo {author} {\bibfnamefont {O.}~\bibnamefont {Reula}},
  \bibinfo {author} {\bibfnamefont {F.}~\bibnamefont {Carrasco}}, \ and\
  \bibinfo {author} {\bibfnamefont {C.}~\bibnamefont {Bederian}},\ }\href
  {\doibase 10.1093/mnras/stac1857} {\bibfield  {journal} {\bibinfo  {journal}
  {Monthly Notices of the Royal Astronomical Society}\ }\textbf {\bibinfo
  {volume} {515}},\ \bibinfo {pages} {1316} (\bibinfo {year}
  {2022})}\BibitemShut {NoStop}%
\bibitem [{\citenamefont {Dexter}\ and\ \citenamefont {Agol}(2009)}]{Acc_1}%
  \BibitemOpen
  \bibfield  {author} {\bibinfo {author} {\bibfnamefont {J.}~\bibnamefont
  {Dexter}}\ and\ \bibinfo {author} {\bibfnamefont {E.}~\bibnamefont {Agol}},\
  }\href {\doibase 10.1088/0004-637X/696/2/1616} {\bibfield  {journal}
  {\bibinfo  {journal} {The Astrophysical Journal}\ }\textbf {\bibinfo {volume}
  {696}},\ \bibinfo {pages} {1616} (\bibinfo {year} {2009})}\BibitemShut
  {NoStop}%
\bibitem [{\citenamefont {Marck}(1996)}]{Acc_2}%
  \BibitemOpen
  \bibfield  {author} {\bibinfo {author} {\bibfnamefont {J.-A.}\ \bibnamefont
  {Marck}},\ }\href {\doibase 10.1088/0264-9381/13/3/007} {\bibfield  {journal}
  {\bibinfo  {journal} {Classical and Quantum Gravity}\ }\textbf {\bibinfo
  {volume} {13}},\ \bibinfo {pages} {393} (\bibinfo {year} {1996})}\BibitemShut
  {NoStop}%
\bibitem [{\citenamefont {Li}\ and\ \citenamefont {Bambi}(2014)}]{Li_14}%
  \BibitemOpen
  \bibfield  {author} {\bibinfo {author} {\bibfnamefont {Z.}~\bibnamefont
  {Li}}\ and\ \bibinfo {author} {\bibfnamefont {C.}~\bibnamefont {Bambi}},\
  }\href {\doibase 10.1088/1475-7516/2014/01/041} {\bibfield  {journal}
  {\bibinfo  {journal} {Journal of Cosmology and Astroparticle Physics}\
  }\textbf {\bibinfo {volume} {2014}},\ \bibinfo {pages} {041} (\bibinfo {year}
  {2014})}\BibitemShut {NoStop}%
\bibitem [{\citenamefont {Wei}\ \emph {et~al.}(2019)\citenamefont {Wei},
  \citenamefont {Zou}, \citenamefont {Liu},\ and\ \citenamefont
  {Mann}}]{Wei_19}%
  \BibitemOpen
  \bibfield  {author} {\bibinfo {author} {\bibfnamefont {S.-W.}\ \bibnamefont
  {Wei}}, \bibinfo {author} {\bibfnamefont {Y.-C.}\ \bibnamefont {Zou}},
  \bibinfo {author} {\bibfnamefont {Y.-X.}\ \bibnamefont {Liu}}, \ and\
  \bibinfo {author} {\bibfnamefont {R.~B.}\ \bibnamefont {Mann}},\ }\href
  {\doibase 10.1088/1475-7516/2019/08/030} {\bibfield  {journal} {\bibinfo
  {journal} {Journal of Cosmology and Astroparticle Physics}\ }\textbf
  {\bibinfo {volume} {2019}},\ \bibinfo {pages} {030} (\bibinfo {year}
  {2019})}\BibitemShut {NoStop}%
\bibitem [{\citenamefont {Cunha}\ \emph {et~al.}(2016)\citenamefont {Cunha},
  \citenamefont {Herdeiro}, \citenamefont {Radu},\ and\ \citenamefont
  {R\'{u}narsson}}]{Cunha_16}%
  \BibitemOpen
  \bibfield  {author} {\bibinfo {author} {\bibfnamefont {P.~V.~P.}\
  \bibnamefont {Cunha}}, \bibinfo {author} {\bibfnamefont {C.~A.~R.}\
  \bibnamefont {Herdeiro}}, \bibinfo {author} {\bibfnamefont {E.}~\bibnamefont
  {Radu}}, \ and\ \bibinfo {author} {\bibfnamefont {H.~F.}\ \bibnamefont
  {R\'{u}narsson}},\ }\href {\doibase 10.1142/S0218271816410212} {\bibfield
  {journal} {\bibinfo  {journal} {International Journal of Modern Physics D}\
  }\textbf {\bibinfo {volume} {25}},\ \bibinfo {pages} {1641021} (\bibinfo
  {year} {2016})}\BibitemShut {NoStop}%
\bibitem [{\citenamefont {de~Vries}(2000)}]{Vries_00}%
  \BibitemOpen
  \bibfield  {author} {\bibinfo {author} {\bibfnamefont {A.}~\bibnamefont
  {de~Vries}},\ }\href {\doibase 10.1088/0264-9381/17/1/309} {\bibfield
  {journal} {\bibinfo  {journal} {Classical and Quantum Gravity}\ }\textbf
  {\bibinfo {volume} {17}},\ \bibinfo {pages} {123} (\bibinfo {year}
  {2000})}\BibitemShut {NoStop}%
\bibitem [{\citenamefont {Abdujabbarov}\ \emph {et~al.}(2016)\citenamefont
  {Abdujabbarov}, \citenamefont {Amir}, \citenamefont {Ahmedov},\ and\
  \citenamefont {Ghosh}}]{Ahmadjon_16}%
  \BibitemOpen
  \bibfield  {author} {\bibinfo {author} {\bibfnamefont {A.}~\bibnamefont
  {Abdujabbarov}}, \bibinfo {author} {\bibfnamefont {M.}~\bibnamefont {Amir}},
  \bibinfo {author} {\bibfnamefont {B.}~\bibnamefont {Ahmedov}}, \ and\
  \bibinfo {author} {\bibfnamefont {S.~G.}\ \bibnamefont {Ghosh}},\ }\href
  {\doibase 10.1103/PhysRevD.93.104004} {\bibfield  {journal} {\bibinfo
  {journal} {Phys. Rev. D}\ }\textbf {\bibinfo {volume} {93}},\ \bibinfo
  {pages} {104004} (\bibinfo {year} {2016})}\BibitemShut {NoStop}%
\bibitem [{\citenamefont {Li}\ \emph {et~al.}(2021)\citenamefont {Li},
  \citenamefont {Abdujabbarov},\ and\ \citenamefont {Han}}]{Li_21}%
  \BibitemOpen
  \bibfield  {author} {\bibinfo {author} {\bibfnamefont {S.}~\bibnamefont
  {Li}}, \bibinfo {author} {\bibfnamefont {A.~A.}\ \bibnamefont
  {Abdujabbarov}}, \ and\ \bibinfo {author} {\bibfnamefont {W.-B.}\
  \bibnamefont {Han}},\ }\href {\doibase 10.1140/epjc/s10052-021-09445-6}
  {\bibfield  {journal} {\bibinfo  {journal} {The European Physical Journal C}\
  }\textbf {\bibinfo {volume} {81}},\ \bibinfo {pages} {649} (\bibinfo {year}
  {2021})}\BibitemShut {NoStop}%
\bibitem [{\citenamefont {Li}\ \emph {et~al.}(2022)\citenamefont {Li},
  \citenamefont {Mirzaev}, \citenamefont {Abdujabbarov}, \citenamefont
  {Malafarina}, \citenamefont {Ahmedov},\ and\ \citenamefont {Han}}]{Li_22}%
  \BibitemOpen
  \bibfield  {author} {\bibinfo {author} {\bibfnamefont {S.}~\bibnamefont
  {Li}}, \bibinfo {author} {\bibfnamefont {T.}~\bibnamefont {Mirzaev}},
  \bibinfo {author} {\bibfnamefont {A.~A.}\ \bibnamefont {Abdujabbarov}},
  \bibinfo {author} {\bibfnamefont {D.}~\bibnamefont {Malafarina}}, \bibinfo
  {author} {\bibfnamefont {B.}~\bibnamefont {Ahmedov}}, \ and\ \bibinfo
  {author} {\bibfnamefont {W.-B.}\ \bibnamefont {Han}},\ }\href {\doibase
  10.1103/PhysRevD.106.084041} {\bibfield  {journal} {\bibinfo  {journal}
  {Phys. Rev. D}\ }\textbf {\bibinfo {volume} {106}},\ \bibinfo {pages}
  {084041} (\bibinfo {year} {2022})}\BibitemShut {NoStop}%
\bibitem [{\citenamefont {Jusufi}\ \emph {et~al.}(2020)\citenamefont {Jusufi},
  \citenamefont {Jamil}, \citenamefont {Chakrabarty}, \citenamefont {Wu},
  \citenamefont {Bambi},\ and\ \citenamefont {Wang}}]{Jusufi_20}%
  \BibitemOpen
  \bibfield  {author} {\bibinfo {author} {\bibfnamefont {K.}~\bibnamefont
  {Jusufi}}, \bibinfo {author} {\bibfnamefont {M.}~\bibnamefont {Jamil}},
  \bibinfo {author} {\bibfnamefont {H.}~\bibnamefont {Chakrabarty}}, \bibinfo
  {author} {\bibfnamefont {Q.}~\bibnamefont {Wu}}, \bibinfo {author}
  {\bibfnamefont {C.}~\bibnamefont {Bambi}}, \ and\ \bibinfo {author}
  {\bibfnamefont {A.}~\bibnamefont {Wang}},\ }\href {\doibase
  10.1103/PhysRevD.101.044035} {\bibfield  {journal} {\bibinfo  {journal}
  {Phys. Rev. D}\ }\textbf {\bibinfo {volume} {101}},\ \bibinfo {pages}
  {044035} (\bibinfo {year} {2020})}\BibitemShut {NoStop}%
\bibitem [{\citenamefont {P\"urrer}(2016)}]{Model_Nor_1}%
  \BibitemOpen
  \bibfield  {author} {\bibinfo {author} {\bibfnamefont {M.}~\bibnamefont
  {P\"urrer}},\ }\href {\doibase 10.1103/PhysRevD.93.064041} {\bibfield
  {journal} {\bibinfo  {journal} {Phys. Rev. D}\ }\textbf {\bibinfo {volume}
  {93}},\ \bibinfo {pages} {064041} (\bibinfo {year} {2016})}\BibitemShut
  {NoStop}%
\bibitem [{\citenamefont {Pan}\ \emph {et~al.}(2008)\citenamefont {Pan},
  \citenamefont {Buonanno}, \citenamefont {Baker}, \citenamefont {Centrella},
  \citenamefont {Kelly}, \citenamefont {McWilliams}, \citenamefont
  {Pretorius},\ and\ \citenamefont {van Meter}}]{Model_Nor_2}%
  \BibitemOpen
  \bibfield  {author} {\bibinfo {author} {\bibfnamefont {Y.}~\bibnamefont
  {Pan}}, \bibinfo {author} {\bibfnamefont {A.}~\bibnamefont {Buonanno}},
  \bibinfo {author} {\bibfnamefont {J.~G.}\ \bibnamefont {Baker}}, \bibinfo
  {author} {\bibfnamefont {J.}~\bibnamefont {Centrella}}, \bibinfo {author}
  {\bibfnamefont {B.~J.}\ \bibnamefont {Kelly}}, \bibinfo {author}
  {\bibfnamefont {S.~T.}\ \bibnamefont {McWilliams}}, \bibinfo {author}
  {\bibfnamefont {F.}~\bibnamefont {Pretorius}}, \ and\ \bibinfo {author}
  {\bibfnamefont {J.~R.}\ \bibnamefont {van Meter}},\ }\href {\doibase
  10.1103/PhysRevD.77.024014} {\bibfield  {journal} {\bibinfo  {journal} {Phys.
  Rev. D}\ }\textbf {\bibinfo {volume} {77}},\ \bibinfo {pages} {024014}
  (\bibinfo {year} {2008})}\BibitemShut {NoStop}%
\bibitem [{\citenamefont {Khan}\ \emph
  {et~al.}(2016{\natexlab{a}})\citenamefont {Khan}, \citenamefont {Husa},
  \citenamefont {Hannam}, \citenamefont {Ohme}, \citenamefont {P\"urrer},
  \citenamefont {Forteza},\ and\ \citenamefont {Boh\'e}}]{Model_Nor_3}%
  \BibitemOpen
  \bibfield  {author} {\bibinfo {author} {\bibfnamefont {S.}~\bibnamefont
  {Khan}}, \bibinfo {author} {\bibfnamefont {S.}~\bibnamefont {Husa}}, \bibinfo
  {author} {\bibfnamefont {M.}~\bibnamefont {Hannam}}, \bibinfo {author}
  {\bibfnamefont {F.}~\bibnamefont {Ohme}}, \bibinfo {author} {\bibfnamefont
  {M.}~\bibnamefont {P\"urrer}}, \bibinfo {author} {\bibfnamefont {X.~J.}\
  \bibnamefont {Forteza}}, \ and\ \bibinfo {author} {\bibfnamefont
  {A.}~\bibnamefont {Boh\'e}},\ }\href {\doibase 10.1103/PhysRevD.93.044007}
  {\bibfield  {journal} {\bibinfo  {journal} {Phys. Rev. D}\ }\textbf {\bibinfo
  {volume} {93}},\ \bibinfo {pages} {044007} (\bibinfo {year}
  {2016}{\natexlab{a}})}\BibitemShut {NoStop}%
\bibitem [{\citenamefont {Husa}\ \emph {et~al.}(2016)\citenamefont {Husa},
  \citenamefont {Khan}, \citenamefont {Hannam}, \citenamefont {P\"urrer},
  \citenamefont {Ohme}, \citenamefont {Forteza},\ and\ \citenamefont
  {Boh\'e}}]{Model_Nor_4}%
  \BibitemOpen
  \bibfield  {author} {\bibinfo {author} {\bibfnamefont {S.}~\bibnamefont
  {Husa}}, \bibinfo {author} {\bibfnamefont {S.}~\bibnamefont {Khan}}, \bibinfo
  {author} {\bibfnamefont {M.}~\bibnamefont {Hannam}}, \bibinfo {author}
  {\bibfnamefont {M.}~\bibnamefont {P\"urrer}}, \bibinfo {author}
  {\bibfnamefont {F.}~\bibnamefont {Ohme}}, \bibinfo {author} {\bibfnamefont
  {X.~J.}\ \bibnamefont {Forteza}}, \ and\ \bibinfo {author} {\bibfnamefont
  {A.}~\bibnamefont {Boh\'e}},\ }\href {\doibase 10.1103/PhysRevD.93.044006}
  {\bibfield  {journal} {\bibinfo  {journal} {Phys. Rev. D}\ }\textbf {\bibinfo
  {volume} {93}},\ \bibinfo {pages} {044006} (\bibinfo {year}
  {2016})}\BibitemShut {NoStop}%
\bibitem [{\citenamefont {Buonanno}\ and\ \citenamefont
  {Damour}(2000)}]{Model_Nor_5}%
  \BibitemOpen
  \bibfield  {author} {\bibinfo {author} {\bibfnamefont {A.}~\bibnamefont
  {Buonanno}}\ and\ \bibinfo {author} {\bibfnamefont {T.}~\bibnamefont
  {Damour}},\ }\href {\doibase 10.1103/PhysRevD.62.064015} {\bibfield
  {journal} {\bibinfo  {journal} {Phys. Rev. D}\ }\textbf {\bibinfo {volume}
  {62}},\ \bibinfo {pages} {064015} (\bibinfo {year} {2000})}\BibitemShut
  {NoStop}%
\bibitem [{\citenamefont {Barausse}\ and\ \citenamefont
  {Buonanno}(2010)}]{Model_Nor_6}%
  \BibitemOpen
  \bibfield  {author} {\bibinfo {author} {\bibfnamefont {E.}~\bibnamefont
  {Barausse}}\ and\ \bibinfo {author} {\bibfnamefont {A.}~\bibnamefont
  {Buonanno}},\ }\href {\doibase 10.1103/PhysRevD.81.084024} {\bibfield
  {journal} {\bibinfo  {journal} {Phys. Rev. D}\ }\textbf {\bibinfo {volume}
  {81}},\ \bibinfo {pages} {084024} (\bibinfo {year} {2010})}\BibitemShut
  {NoStop}%
\bibitem [{\citenamefont {Buonanno}\ \emph {et~al.}(2009)\citenamefont
  {Buonanno}, \citenamefont {Pan}, \citenamefont {Pfeiffer}, \citenamefont
  {Scheel}, \citenamefont {Buchman},\ and\ \citenamefont
  {Kidder}}]{Model_Nor_7}%
  \BibitemOpen
  \bibfield  {author} {\bibinfo {author} {\bibfnamefont {A.}~\bibnamefont
  {Buonanno}}, \bibinfo {author} {\bibfnamefont {Y.}~\bibnamefont {Pan}},
  \bibinfo {author} {\bibfnamefont {H.~P.}\ \bibnamefont {Pfeiffer}}, \bibinfo
  {author} {\bibfnamefont {M.~A.}\ \bibnamefont {Scheel}}, \bibinfo {author}
  {\bibfnamefont {L.~T.}\ \bibnamefont {Buchman}}, \ and\ \bibinfo {author}
  {\bibfnamefont {L.~E.}\ \bibnamefont {Kidder}},\ }\href {\doibase
  10.1103/PhysRevD.79.124028} {\bibfield  {journal} {\bibinfo  {journal} {Phys.
  Rev. D}\ }\textbf {\bibinfo {volume} {79}},\ \bibinfo {pages} {124028}
  (\bibinfo {year} {2009})}\BibitemShut {NoStop}%
\bibitem [{\citenamefont {Damour}\ \emph {et~al.}(2008)\citenamefont {Damour},
  \citenamefont {Nagar}, \citenamefont {Hannam}, \citenamefont {Husa},\ and\
  \citenamefont {Br\"ugmann}}]{Model_Nor_8}%
  \BibitemOpen
  \bibfield  {author} {\bibinfo {author} {\bibfnamefont {T.}~\bibnamefont
  {Damour}}, \bibinfo {author} {\bibfnamefont {A.}~\bibnamefont {Nagar}},
  \bibinfo {author} {\bibfnamefont {M.}~\bibnamefont {Hannam}}, \bibinfo
  {author} {\bibfnamefont {S.}~\bibnamefont {Husa}}, \ and\ \bibinfo {author}
  {\bibfnamefont {B.}~\bibnamefont {Br\"ugmann}},\ }\href {\doibase
  10.1103/PhysRevD.78.044039} {\bibfield  {journal} {\bibinfo  {journal} {Phys.
  Rev. D}\ }\textbf {\bibinfo {volume} {78}},\ \bibinfo {pages} {044039}
  (\bibinfo {year} {2008})}\BibitemShut {NoStop}%
\bibitem [{\citenamefont {Pan}\ \emph {et~al.}(2010)\citenamefont {Pan},
  \citenamefont {Buonanno}, \citenamefont {Buchman}, \citenamefont {Chu},
  \citenamefont {Kidder}, \citenamefont {Pfeiffer},\ and\ \citenamefont
  {Scheel}}]{Model_Nor_9}%
  \BibitemOpen
  \bibfield  {author} {\bibinfo {author} {\bibfnamefont {Y.}~\bibnamefont
  {Pan}}, \bibinfo {author} {\bibfnamefont {A.}~\bibnamefont {Buonanno}},
  \bibinfo {author} {\bibfnamefont {L.~T.}\ \bibnamefont {Buchman}}, \bibinfo
  {author} {\bibfnamefont {T.}~\bibnamefont {Chu}}, \bibinfo {author}
  {\bibfnamefont {L.~E.}\ \bibnamefont {Kidder}}, \bibinfo {author}
  {\bibfnamefont {H.~P.}\ \bibnamefont {Pfeiffer}}, \ and\ \bibinfo {author}
  {\bibfnamefont {M.~A.}\ \bibnamefont {Scheel}},\ }\href {\doibase
  10.1103/PhysRevD.81.084041} {\bibfield  {journal} {\bibinfo  {journal} {Phys.
  Rev. D}\ }\textbf {\bibinfo {volume} {81}},\ \bibinfo {pages} {084041}
  (\bibinfo {year} {2010})}\BibitemShut {NoStop}%
\bibitem [{\citenamefont {Hannam}\ \emph {et~al.}(2014)\citenamefont {Hannam},
  \citenamefont {Schmidt}, \citenamefont {Boh\'e}, \citenamefont {Haegel},
  \citenamefont {Husa}, \citenamefont {Ohme}, \citenamefont {Pratten},\ and\
  \citenamefont {P\"urrer}}]{Model_Nor_10}%
  \BibitemOpen
  \bibfield  {author} {\bibinfo {author} {\bibfnamefont {M.}~\bibnamefont
  {Hannam}}, \bibinfo {author} {\bibfnamefont {P.}~\bibnamefont {Schmidt}},
  \bibinfo {author} {\bibfnamefont {A.}~\bibnamefont {Boh\'e}}, \bibinfo
  {author} {\bibfnamefont {L.}~\bibnamefont {Haegel}}, \bibinfo {author}
  {\bibfnamefont {S.}~\bibnamefont {Husa}}, \bibinfo {author} {\bibfnamefont
  {F.}~\bibnamefont {Ohme}}, \bibinfo {author} {\bibfnamefont {G.}~\bibnamefont
  {Pratten}}, \ and\ \bibinfo {author} {\bibfnamefont {M.}~\bibnamefont
  {P\"urrer}},\ }\href {\doibase 10.1103/PhysRevLett.113.151101} {\bibfield
  {journal} {\bibinfo  {journal} {Phys. Rev. Lett.}\ }\textbf {\bibinfo
  {volume} {113}},\ \bibinfo {pages} {151101} (\bibinfo {year}
  {2014})}\BibitemShut {NoStop}%
\bibitem [{\citenamefont {Giesler}\ \emph {et~al.}(2019)\citenamefont
  {Giesler}, \citenamefont {Isi}, \citenamefont {Scheel},\ and\ \citenamefont
  {Teukolsky}}]{Giesler_19}%
  \BibitemOpen
  \bibfield  {author} {\bibinfo {author} {\bibfnamefont {M.}~\bibnamefont
  {Giesler}}, \bibinfo {author} {\bibfnamefont {M.}~\bibnamefont {Isi}},
  \bibinfo {author} {\bibfnamefont {M.~A.}\ \bibnamefont {Scheel}}, \ and\
  \bibinfo {author} {\bibfnamefont {S.~A.}\ \bibnamefont {Teukolsky}},\ }\href
  {\doibase 10.1103/PhysRevX.9.041060} {\bibfield  {journal} {\bibinfo
  {journal} {Phys. Rev. X}\ }\textbf {\bibinfo {volume} {9}},\ \bibinfo {pages}
  {041060} (\bibinfo {year} {2019})}\BibitemShut {NoStop}%
\bibitem [{\citenamefont {Ota}\ and\ \citenamefont {Chirenti}(2020)}]{Iara_20}%
  \BibitemOpen
  \bibfield  {author} {\bibinfo {author} {\bibfnamefont {I.}~\bibnamefont
  {Ota}}\ and\ \bibinfo {author} {\bibfnamefont {C.}~\bibnamefont {Chirenti}},\
  }\href {\doibase 10.1103/PhysRevD.101.104005} {\bibfield  {journal} {\bibinfo
   {journal} {Phys. Rev. D}\ }\textbf {\bibinfo {volume} {101}},\ \bibinfo
  {pages} {104005} (\bibinfo {year} {2020})}\BibitemShut {NoStop}%
\bibitem [{\citenamefont {Dhani}(2021)}]{Arnab_21}%
  \BibitemOpen
  \bibfield  {author} {\bibinfo {author} {\bibfnamefont {A.}~\bibnamefont
  {Dhani}},\ }\href {\doibase 10.1103/PhysRevD.103.104048} {\bibfield
  {journal} {\bibinfo  {journal} {Phys. Rev. D}\ }\textbf {\bibinfo {volume}
  {103}},\ \bibinfo {pages} {104048} (\bibinfo {year} {2021})}\BibitemShut
  {NoStop}%
\bibitem [{\citenamefont {Yang}\ \emph {et~al.}(2012)\citenamefont {Yang},
  \citenamefont {Nichols}, \citenamefont {Zhang}, \citenamefont {Zimmerman},
  \citenamefont {Zhang},\ and\ \citenamefont {Chen}}]{Yang_12}%
  \BibitemOpen
  \bibfield  {author} {\bibinfo {author} {\bibfnamefont {H.}~\bibnamefont
  {Yang}}, \bibinfo {author} {\bibfnamefont {D.~A.}\ \bibnamefont {Nichols}},
  \bibinfo {author} {\bibfnamefont {F.}~\bibnamefont {Zhang}}, \bibinfo
  {author} {\bibfnamefont {A.}~\bibnamefont {Zimmerman}}, \bibinfo {author}
  {\bibfnamefont {Z.}~\bibnamefont {Zhang}}, \ and\ \bibinfo {author}
  {\bibfnamefont {Y.}~\bibnamefont {Chen}},\ }\href {\doibase
  10.1103/PhysRevD.86.104006} {\bibfield  {journal} {\bibinfo  {journal} {Phys.
  Rev. D}\ }\textbf {\bibinfo {volume} {86}},\ \bibinfo {pages} {104006}
  (\bibinfo {year} {2012})}\BibitemShut {NoStop}%
\bibitem [{\citenamefont {Li}\ and\ \citenamefont {Han}(2022)}]{Psi_22}%
  \BibitemOpen
  \bibfield  {author} {\bibinfo {author} {\bibfnamefont {S.}~\bibnamefont
  {Li}}\ and\ \bibinfo {author} {\bibfnamefont {W.-B.}\ \bibnamefont {Han}},\
  }\href {\doibase 10.1103/PhysRevD.106.104013} {\bibfield  {journal} {\bibinfo
   {journal} {Phys. Rev. D}\ }\textbf {\bibinfo {volume} {106}},\ \bibinfo
  {pages} {104013} (\bibinfo {year} {2022})}\BibitemShut {NoStop}%
\bibitem [{\citenamefont {Collins}\ and\ \citenamefont
  {Hughes}(2004)}]{Bumpy_BH_1}%
  \BibitemOpen
  \bibfield  {author} {\bibinfo {author} {\bibfnamefont {N.~A.}\ \bibnamefont
  {Collins}}\ and\ \bibinfo {author} {\bibfnamefont {S.~A.}\ \bibnamefont
  {Hughes}},\ }\href {\doibase 10.1103/PhysRevD.69.124022} {\bibfield
  {journal} {\bibinfo  {journal} {Phys. Rev. D}\ }\textbf {\bibinfo {volume}
  {69}},\ \bibinfo {pages} {124022} (\bibinfo {year} {2004})}\BibitemShut
  {NoStop}%
\bibitem [{\citenamefont {Vigeland}\ and\ \citenamefont
  {Hughes}(2010)}]{Bumpy_BH_2}%
  \BibitemOpen
  \bibfield  {author} {\bibinfo {author} {\bibfnamefont {S.~J.}\ \bibnamefont
  {Vigeland}}\ and\ \bibinfo {author} {\bibfnamefont {S.~A.}\ \bibnamefont
  {Hughes}},\ }\href {\doibase 10.1103/PhysRevD.81.024030} {\bibfield
  {journal} {\bibinfo  {journal} {Phys. Rev. D}\ }\textbf {\bibinfo {volume}
  {81}},\ \bibinfo {pages} {024030} (\bibinfo {year} {2010})}\BibitemShut
  {NoStop}%
\bibitem [{\citenamefont {Shashank}\ and\ \citenamefont
  {Bambi}(2022{\natexlab{a}})}]{KRZ_ins}%
  \BibitemOpen
  \bibfield  {author} {\bibinfo {author} {\bibfnamefont {S.}~\bibnamefont
  {Shashank}}\ and\ \bibinfo {author} {\bibfnamefont {C.}~\bibnamefont
  {Bambi}},\ }\href {\doibase 10.1103/PhysRevD.105.104004} {\bibfield
  {journal} {\bibinfo  {journal} {Phys. Rev. D}\ }\textbf {\bibinfo {volume}
  {105}},\ \bibinfo {pages} {104004} (\bibinfo {year}
  {2022}{\natexlab{a}})}\BibitemShut {NoStop}%

\bibitem [{\citenamefont {Khan}\ \emph
  {et~al.}(2016{\natexlab{b}})\citenamefont {Khan}, \citenamefont {Husa},
  \citenamefont {Hannam}, \citenamefont {Ohme}, \citenamefont {P\"urrer},
  \citenamefont {Forteza},\ and\ \citenamefont {Boh\'e}}]{phenomD}%
  \BibitemOpen
  \bibfield  {author} {\bibinfo {author} {\bibfnamefont {S.}~\bibnamefont
  {Khan}}, \bibinfo {author} {\bibfnamefont {S.}~\bibnamefont {Husa}}, \bibinfo
  {author} {\bibfnamefont {M.}~\bibnamefont {Hannam}}, \bibinfo {author}
  {\bibfnamefont {F.}~\bibnamefont {Ohme}}, \bibinfo {author} {\bibfnamefont
  {M.}~\bibnamefont {P\"urrer}}, \bibinfo {author} {\bibfnamefont {X.~J.}\
  \bibnamefont {Forteza}}, \ and\ \bibinfo {author} {\bibfnamefont
  {A.}~\bibnamefont {Boh\'e}},\ }\href {\doibase 10.1103/PhysRevD.93.044007}
  {\bibfield  {journal} {\bibinfo  {journal} {Phys. Rev. D}\ }\textbf {\bibinfo
  {volume} {93}},\ \bibinfo {pages} {044007} (\bibinfo {year}
  {2016}{\natexlab{b}})}\BibitemShut {NoStop}%
\bibitem [{\citenamefont {McWilliams}(2019)}]{McWilliams_19}%
  \BibitemOpen
  \bibfield  {author} {\bibinfo {author} {\bibfnamefont {S.~T.}\ \bibnamefont
  {McWilliams}},\ }\href {\doibase 10.1103/PhysRevLett.122.191102} {\bibfield
  {journal} {\bibinfo  {journal} {Phys. Rev. Lett.}\ }\textbf {\bibinfo
  {volume} {122}},\ \bibinfo {pages} {191102} (\bibinfo {year}
  {2019})}\BibitemShut {NoStop}%
\bibitem [{\citenamefont {Ma}\ \emph {et~al.}(2021)\citenamefont {Ma},
  \citenamefont {Giesler}, \citenamefont {Varma}, \citenamefont {Scheel},\ and\
  \citenamefont {Chen}}]{BOB_b}%
  \BibitemOpen
  \bibfield  {author} {\bibinfo {author} {\bibfnamefont {S.}~\bibnamefont
  {Ma}}, \bibinfo {author} {\bibfnamefont {M.}~\bibnamefont {Giesler}},
  \bibinfo {author} {\bibfnamefont {V.}~\bibnamefont {Varma}}, \bibinfo
  {author} {\bibfnamefont {M.~A.}\ \bibnamefont {Scheel}}, \ and\ \bibinfo
  {author} {\bibfnamefont {Y.}~\bibnamefont {Chen}},\ }\href {\doibase
  10.1103/PhysRevD.104.084003} {\bibfield  {journal} {\bibinfo  {journal}
  {Phys. Rev. D}\ }\textbf {\bibinfo {volume} {104}},\ \bibinfo {pages}
  {084003} (\bibinfo {year} {2021})}\BibitemShut {NoStop}%
\bibitem [{\citenamefont {Harry}\ and\ \citenamefont {(forthe LIGO
  Scientific~Collaboration)}(2010)}]{Harry_2010}%
  \BibitemOpen
  \bibfield  {author} {\bibinfo {author} {\bibfnamefont {G.~M.}\ \bibnamefont
  {Harry}}\ and\ \bibinfo {author} {\bibnamefont {(forthe LIGO
  Scientific~Collaboration)}},\ }\href {\doibase 10.1088/0264-9381/27/8/084006}
  {\bibfield  {journal} {\bibinfo  {journal} {Classical and Quantum Gravity}\
  }\textbf {\bibinfo {volume} {27}},\ \bibinfo {pages} {084006} (\bibinfo
  {year} {2010})}\BibitemShut {NoStop}%
\bibitem [{\citenamefont {Amaro-Seoane}\ \emph {et~al.}(2023)\citenamefont
  {Amaro-Seoane} \emph {et~al.}}]{LISA}%
  \BibitemOpen
  \bibfield  {author} {\bibinfo {author} {\bibfnamefont {P.}~\bibnamefont
  {Amaro-Seoane}} \emph {et~al.},\ }\href {\doibase 10.1007/s41114-022-00041-y}
  {\bibfield  {journal} {\bibinfo  {journal} {Living Reviews in Relativity}\
  }\textbf {\bibinfo {volume} {26}},\ \bibinfo {pages} {2} (\bibinfo {year}
  {2023})}\BibitemShut {NoStop}%
\bibitem [{\citenamefont {Hild}\ \emph {et~al.}(2011)\citenamefont {Hild} \emph
  {et~al.}}]{ET_D}%
  \BibitemOpen
  \bibfield  {author} {\bibinfo {author} {\bibfnamefont {S.}~\bibnamefont
  {Hild}} \emph {et~al.},\ }\href {\doibase 10.1088/0264-9381/28/9/094013}
  {\bibfield  {journal} {\bibinfo  {journal} {Classical and Quantum Gravity}\
  }\textbf {\bibinfo {volume} {28}},\ \bibinfo {pages} {094013} (\bibinfo
  {year} {2011})}\BibitemShut {NoStop}%
\end{thebibliography}

\end{document}